\documentclass[iop,twocolumn]{emulateapj}
\usepackage{epsfig}
\usepackage[caption=false]{subfig}
\usepackage{csquotes}
\usepackage{graphicx}
\usepackage{grffile}
\usepackage{ulem,color}
\usepackage[dvipsnames]{xcolor/xcolor}
\usepackage{dcolumn}
\usepackage{bm}
\usepackage{natbib}
\usepackage[bookmarks=true]{hyperref}
\usepackage{xspace}
\usepackage{amsmath}
\usepackage{lipsum}
\usepackage{xfrac}
\def\lesssim{\mathrel{\hbox{\rlap{\hbox{\lower4pt\hbox{$\sim$}}}\hbox{$<$}}}}
\def\gtrsim{\mathrel{\hbox{\rlap{\hbox{\lower4pt\hbox{$\sim$}}}\hbox{$>$}}}}

\newcommand{\bea}{\begin{eqnarray}}
\newcommand{\eea}{\end{eqnarray}}

\newcommand{\dF}{{^{^*}\!\!F}}

\newcommand{\del}{{\partial}}

\newcommand{\harm}{{\sc Harm3d}\xspace}

\newcommand{\dotm}{$\dot M$}

\newcommand{\pwmhd}{\xspace{\sc PatchworkMHD}}

\newcommand{\harmd}{\xspace{\sc Harm3d}\xspace}

\newcommand{\pmhd}{\xspace{\sc PatchworkMHD}\xspace}

\def\lambdabar{%
\relax
\bgroup
\def\@tempa{\hbox{\raise.73\ht0
\hbox to0pt{\kern.25\wd0\vrule width.5\wd0
height.1pt depth.1pt\hss}\box0}}%
\mathchoice{\setbox0\hbox{$\displaystyle\lambda$}\@tempa}%
{\setbox0\hbox{$\textstyle\lambda$}\@tempa}%
{\setbox0\hbox{$\scriptstyle\lambda$}\@tempa}%
{\setbox0\hbox{$\scriptscriptstyle\lambda$}\@tempa}%
\egroup
}

\slugcomment{Submitted to ApJ.}

\newcommand{\outline}[1]{
	}

\usepackage{enumitem}
\setlist[itemize]{noitemsep} 

\begin{document}

\title{Accretion Onto A Supermassive Black Hole Binary Before Merger}

\author{Mark J. Avara$^{1,2}$, Julian H. Krolik$^3$, Manuela Campanelli$^1$, Scott C. Noble$^4$, Dennis Bowen$^5$, Taeho Ryu$^{3,6}$}

\affil{
  $^1$ Institute of Astronomy, University of Cambridge, Madingley Road, Cambridge CB3 0HA, UK\\
  $^2$ Center for Computational Relativity and Gravitation, Rochester Institute of Technology, Rochester, NY 14623, USA\\
  $^3$ William H. Miller Department of Physics and Astronomy, Johns Hopkins University, Baltimore MD 21218, USA\\
  $^4$ Gravitational Astrophysics Lab, NASA Goddard Space Flight Center, Greenbelt, Maryland 20771, USA \\
  $^5$ X Computational Physics, Los Alamos National Laboratory, P.O. Box 1663, Los Alamos, New Mexico 87545, USA\\
  $^6$ Max Planck Institute for Astrophysics, Karl-Schwarzschild-Str.1, 85748 Garching, Germany}

\email{mjavara@gmail.com}


\begin{abstract}
  While supermassive binary black holes (SMBBHs) inspiral toward merger they may also experience significant accretion of matter from a surrounding disk. To study the dynamics of this system requires simultaneously describing the evolving spacetime and the magnetized plasma. We present the first relativistic calculation simulating two equal-mass, non-spinning black holes as they inspiral from a $20M$ ($G=c=1$) initial separation almost to merger, $\simeq 9M$ ($M$=total binary mass). Our dynamical results imply important observational consequences: for instance, the accretion rate $\dot M$ onto the black holes first decreases and then reaches a plateau, dropping by only a factor of $\sim 3$ despite the rapid inspiral. An estimated bolometric light curve follows the same profile, suggesting some merging SMBBHs may be significantly luminous past the predicted decoupling from the circumbinary disk.   
  The minidisks through which the accretion reaches the black holes are very non-standard: Reynolds, not Maxwell, stresses dominate, and they oscillate between two distinct states. In one part of the cycle, ``sloshing" streams transfer mass from one minidisk to the other through the L1 point at a rate $\sim 0.1\times$ the accretion rate, carrying kinetic energy at a rate that can be as large as the peak minidisk bolometric luminosity. We also discover that episodic accretion drives minidisks with time-varying tilts with respect to the orbital plane. The accretion cycles, energy dissipated by sloshing material, and variable inclination to the observer all contribute to unique cyclical behavior in the light curves of late-time inspiraling SMBBHs. The unsigned poloidal magnetic flux on the black hole event horizon is roughly constant at a dimensionless level $\phi\sim 2-3$, but doubles just before merger; if the black holes had significant spin, this flux indicates the potential for powerful jets with variability driven by binary dynamics, another prediction of potentially unique EM signatures. This simulation is the first to employ our multipatch infrastructure \pwmhd, decreasing computational expense to $\sim 3\%$ 
  of conventional single-grid methods' cost.
\end{abstract}

\keywords{Black hole physics - magnetohydrodynamics - accretion, accretion disks}


\section{Introduction}
\label{sec:introduction}

In the consensus model of cosmology, the population of present-day galaxies grew from the merger of an earlier population of less massive galaxies in a hierarchical fashion \citep{Klein2016,Katz2020}.  Because galaxies today are known to contain supermassive black holes at their centers \citep{KormendyHo2013}, it is natural to think that this chain of mergers frequently brought two supermassive black holes together in newly merged galaxies.  A variety of processes might then bring both black holes to the center of the merged system and close enough to each other to form a gravitationally bound binary \citep{BBR80}.   Global asymmetries in the stellar mass distribution as well as the effects of gas accretion could then tighten the binary's orbit to the point that gravitational wave (GW) radiation drives the two black holes together [see the recent review by \cite{Bogdanovic2022} and references therein], ending with a burst of gravitational wave radiation potentially detectable by future low-frequency GW detectors such as LISA \citep{Mangiagli2020,Baker2019,Kelley2019}, while the approach to merger might be detected by Pulsar Timing Arrays \citep[][and participating experiments]{verbiest2016}.

Although no electromagnetic counterparts to black hole mergers detected by LIGO have been seen, a simple heuristic argument leads to the prediction that such counterparts are much more likely to be associated with LISA observations.  Though accretion onto stellar mass BHs in X-ray binaries may be near or possibly exceed the Eddington luminosity, such high rates of accretion onto stellar mass BBHs would require a source other than massive wind or Roche-lobe overflow from a nearby companion. For isolated black holes and assuming the simplified Bondi solution, the accretion rate of an object of mass $M$ embedded in interstellar gas with density $\rho$ is $\propto M^2 \rho/(v^2 + c_s^2)^{2}$, where $v$ is the bulk speed of the gas relative to the gravitating object and $c_s$ is the thermal speed of its atoms.  Supermassive black holes in galaxy centers are more massive than LIGO black holes by at least a factor $\sim 10^5$ and their nearby interstellar media can be denser by a factor $\sim 10^4$, while $v$ and $c_s$ may not differ by much.  For fixed radiative efficiency in the accretion flow, the luminosity of supermassive binary black holes should therefore be $\sim 10^{14}\times$ that of stellar-mass binary black holes or more.  If this simplified approach retains any validity at all for such binary systems, the contrast in luminosity between stellar mass and supermassive BH binaries should be tremendous.  This argument is also supported by the fact that in several percent of all galaxies in the contemporary Universe, the accretion rate onto a central supermassive black hole is sufficient to power an active galactic nucleus (AGN) with luminosity $\gtrsim 10^{44}$~erg~s$^{-1}$, and this active fraction is considerably higher at redshift $z \sim 2 - 3$ (see \cite{ReinesComastri2016} and references therein).

If at least some merging SMBBHs are also EM bright \citep{Haiman2023}, the scientific value of GW detections can be greatly augmented.   If a LISA source can be identified with a host galaxy, its redshifts can be determined independently, and the stellar mass and evolutionary state of the galaxy can be measured, thereby placing the event in the context of galaxy evolution.  Since it will still be years before such joint detections are possible, it is intriguing to consider the possibility that EM detections of merging SMBBHs might be made {\it without} any GW detection, perhaps even before LISA is launched; with luck, we might even discover systems whose actual merger might be seen during LISA's lifetime.   A broad goal of our line of work is to explore this possibility by identifying unique EM signatures of SMBBH inspiral and mergers.

This motivation is strengthened considering that even with a GW detection, identifying a possible EM counterpart is made difficult by the poor sky-localization capabilities of GW observatories, i.e. extremely large numbers of galaxies could lie within the uncertainty range \citep{Mingarelli2022}.  In the absence of even a crude localization, the problem becomes even harder.  Discovering such a counterpart demands foreknowledge of distinct spectral features or time-dependence to narrow the search.

We have embarked upon a program to determine these EM features.  We begin by building upon what is already known about accretion in binary systems in general.  If the binary mass-ratio $q \equiv M_2/M_1$ is $\gtrsim 0.02$, the disk surrounding the binary is truncated at a radius from the center-of-mass $\simeq (2 - 3) a$, for binary semi-major axis $a$ \citep{ArtymLubow94}.  Although the binary exerts strong positive torques on matter following prograde orbits near the truncation radius \citep{P91}, streams can break off from the circumbinary disk's (CBD) inner edge and convey accreting matter from the outer disk to the binary \citep{ArtymLubow94,MM08,Shi12,Noble12}, ultimately achieving inflow equilibrium \citep{Farris2014,Shi15}.  When the orbit is roughly circular and $q \gtrsim 0.2$ , a ``lump" forms at the inner edge of the CBD and modulates the accretion rate through the inner gap with a frequency $\simeq 0.2\Omega_b$, for binary orbital frequency $\Omega_b$ \citep{Shi12,Noble12,Farris2014,Bowen2018,LopezArmengol2021,Noble21}.  The accretion is then apportioned to two ``minidisks", each surrounding one member of the binary, with the less massive receiving the lion's share \citep{Farris2014,Shi15,Bowen2019,Combi2021}.

This situation is potentially complicated when the evolution rate of the binary (due, for example, to GW radiation) becomes faster than the inflow rate due to ordinary accretion processes in the CBD \citep{MP05}.  Although initial estimates suggested accretion onto the binary would be entirely cut off, detailed simulations [\cite{Noble12}, hereafter ``Noble12" (3D, excised cavity), \cite{Farris2015,DittmannRyanMiller2023,Krauth2023} (2D, cavity included), \cite{Bowen2018,Bowen2019} (3D, most cavity included, short duration), \cite{Combi2021}(3D, most cavity included, short duration)] have supported the argument that it may be reduced by a factor of order unity or even less, but not suppressed altogether.

This paper focuses on magnetized gas dynamics while an equal-mass binary surrounded  by  a coplanar circumbinary  disk inspirals from a separation of $20M$ to $9M$. Even without a prescription for full radiation transport, the results provide insight into EM emission by mapping how much gas is where and what happens to it. We highlight distinct observational signatures related to: the bolometric lightcurves of the CBD and each minidisk, the power in relativistic jets, and the radiation from shocks occurring when ``sloshing" streams strike a minidisk.
We also estimate the gas mass residing near the binary immediately before merger, the single parameter most important to estimating the energy released in photons during the actual merger \citep{Krolik10}. 
These results are key elements for prediction of population statistics of EM counterparts to SMBBH mergers.

Our methods are also novel.
We start with conditions pre-equilibrated to the binary, and evolve them in fully global 3D-GRMHD.  In addition,
we introduce in this paper a conceptually new approach to simulating gas flow around a binary: a multipatch method, in which separate programs compute the fluid's evolution in different spatial portions of the system.  These independent ``patches" all have their own grids and exchange boundary condition data.  A preliminary version of this method applicable to hydrodynamic problems, called {\it Patchwork}, was described in \cite{Shiokawa2018}; here we introduce \pmhd, an extension of \textit{Patchwork} with new methodology for MHD, as well as a number of design and efficiency improvements. In the Appendix we introduce specifics related to the evolution of magnetic fields in \pwmhd+\harmd (for further details and tests, see Avara et~al., in preparation). Use of the multipatch method avoids the problems created by the singularity at the symmetry axis of polar coordinates, and, in so doing, permits time-steps long enough to make this simulation feasible.

In section \S\ref{sec:setup} we give details on the numerical methodology and physical setup, including a summary of the key developments of the \pmhd code that enabled our fiducial runs. We then describe in detail our fiducial simulation, first focusing on the dynamical matter evolution of the system in \S\ref{sec:hydroresults}, and then the magnetic evolution in \S\ref{sec:magnetic}. Finally, in \S\ref{sec:discussion} we discuss our findings, make comparisons with other studies, and conclude in \S\ref{sec:conc}.

\section{Simulation Methodology}
\label{sec:setup}

\subsection{The basic equations}

To determine the evolution of gas in such a system, we must solve the equations of general relativistic MHD, including time-dependence in the metric.   The finite volume MHD code we use in combination with \pwmhd, \harm \citep{Noble09}, poses both the fluid equations and the Faraday equation in a conservative framework:
\begin{equation}
\partial_t {\mathbf U} ({\mathbf P}) = - \partial_i {\mathbf F}^i({\mathbf P}) + {\mathbf S}({\mathbf P}),
\end{equation}
where the vectors ${\mathbf P}$, ${\mathbf U}$, ${\mathbf F}$, and ${\mathbf S}$ represent the primitive variables, the conserved variables, the fluxes and the source terms, respectively.
$\mathbf{P}$ and the three vector functions depending on it are:
\begin{eqnarray}
{\mathbf P} &=& \left[\rho, \epsilon, u^\mu, b^\mu \right],\\
{\mathbf U} &=& \sqrt{-g} \left[ \rho u^t, T^t_t + \rho u^t, T^t_j, B^k\right]^T,\\
{\mathbf F} &=& \sqrt{-g} \left[ \rho u^i, T^i_t + \rho u^i, T^i_j, (b^i u^k - b^k u^i)   \right]^T,\\
{\mathbf S} &=& \sqrt{-g} \left[0, T^\kappa_\lambda \Gamma^\lambda_{t\kappa} - {\cal F}_t, T^\kappa_\lambda \Gamma^\lambda_{j \kappa} - {\cal F}_j,0 \right]^T  .
\end{eqnarray} 
Here $g$ is the metric determinant, $\rho$ is the proper (i.e., fluid rest-frame) mass density, $\epsilon$ is the specific internal energy density, $u^\mu$ is the fluid 4-velocity, $T^\mu_\nu$ is the stress-energy tensor, $\Gamma^\kappa_{\lambda\alpha}$ is the affine connection, $b^\alpha$ is the magnetic 4-vector $\dF^{\beta \alpha} u_\beta$, and $\dF^{\alpha \beta}$ is the dual EM field tensor in which $B^i = \dF^{it}$.  We write the stress-energy tensor as
\begin{equation}
T^\mu_\nu = (\rho h + b^2) u^\mu u_\nu + (p + b^2/2) \delta^\mu_\nu - b^\mu b_\nu,
\end{equation}
where $h \equiv 1 + (\epsilon + p)/\rho$ is the specific enthalpy, $p$ is the gas pressure, and $b^2 = b^\alpha b_\alpha$.
The fluid equations are closed by an equation of state $p = \left(\gamma - 1\right) \rho \epsilon$ with $\gamma=5/3$.

The vector  ${\cal F}_\mu \equiv {\cal L}u_\mu$  represents the 4-momentum lost from a fluid element by radiative cooling.  The magnitude of ${\cal L}$ is determined by comparing the local entropy proxy $S \equiv p/\rho^\gamma$ to a target value $S_*$=0.01:
\begin{equation}
{\cal L} = \begin{cases} (\rho \epsilon/t_{\rm orb})(S/S_* - 1)^{1/2} & S > S_* \\
                     0 & S \leq S_* .\end{cases}
\end{equation}
Here $t_{\rm orb}$ is the period of a circular orbit around the nearest black hole when the distance to that black hole is $< 0.45a$.  When the distance to the center-of-mass is $>1.5a$, it is the period of a circular orbit whose radius is the distance to the center-of-mass.  Everywhere else, it is the period of a circular orbit around the center-of-mass at radius $1.5a$. The cooling is also set to zero wherever the gas is unbound according to the criterion $u_t (\rho h + b^2) < -\rho$ (we adopt a metric signature - + + +).  This cooling rate is designed so as to radiate away nearly all the heat generated by dissipative processes.  It is set to zero in unbound material because this material tends to be unphysically hot; including its radiation in the system's luminosity can therefore be misleading, though we test this choice in the SMBBH context by also simulating with cooling all gas in the system. 

\harm \citep{Noble06,Noble09,Noble12} solves the discretized fluid conservation equations by a finite-volume method utilizing a Lax-Friedrichs Riemann solver.  It solves the Faraday equation by the constrained transport algorithm FluxCT \citep{Toth2000}.  

Although we do not solve the full set of Einstein Field Equations to determine the time-dependent spacetime, we use an excellent approximation to their solution.  To construct this approximation, we asymptotically match Schwarzschild spacetimes around each black hole to a high-order post-Newtonian expansion (2.5PN), while the orbital evolution is calculated to even higher-order, 3.5PN \citep{Noble12,Mundim2014,Ireland2016}.

\subsection{The multipatch method}

\begin{figure*}[htp]
    \centering
	    \subfloat{\includegraphics[trim={1.cm 0cm .5cm 0cm},clip,width=.45\textwidth]{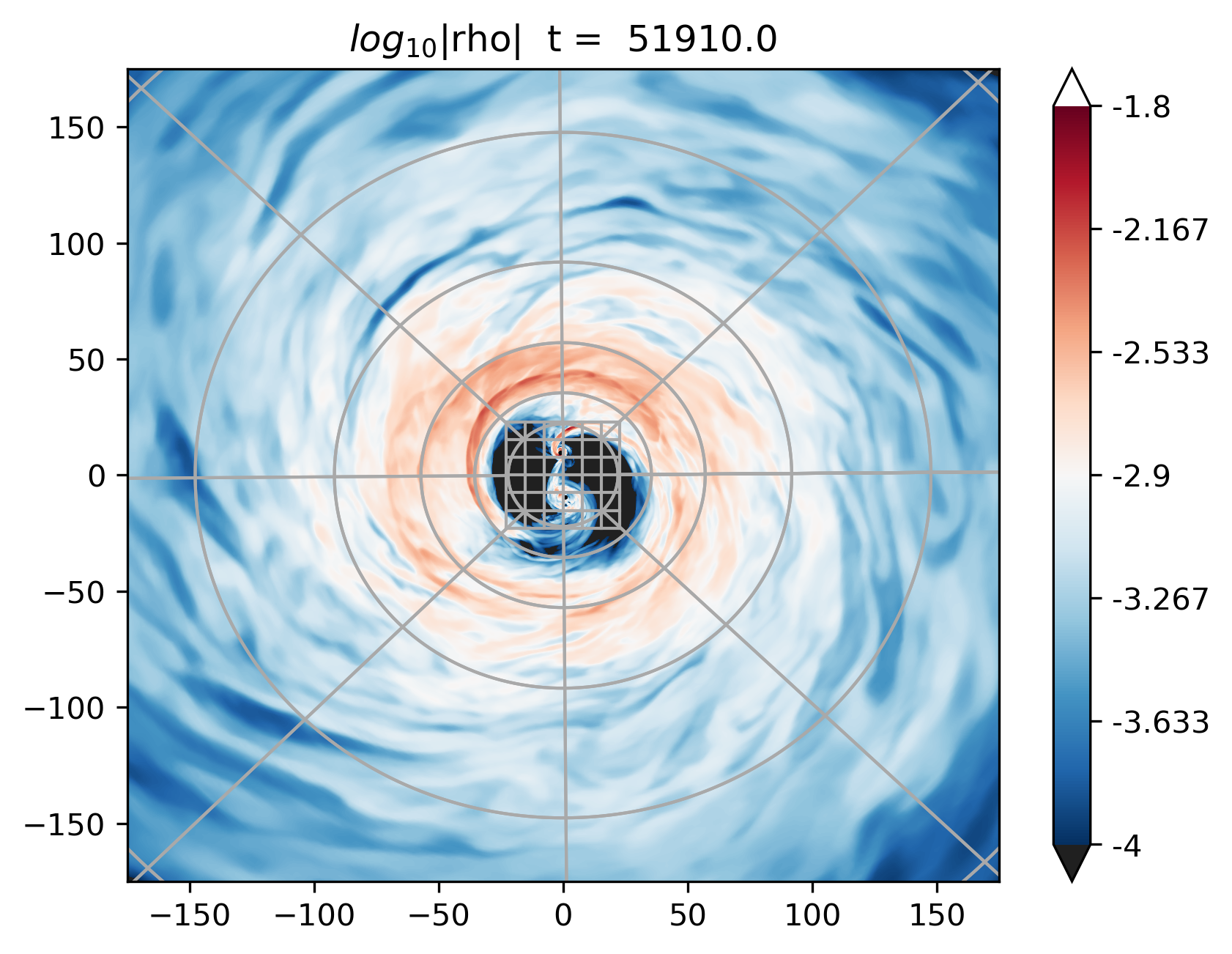} }
	        \quad               
	    \subfloat{\includegraphics[trim={1.cm 0cm 0.25cm 0cm},clip,width=.45\textwidth]{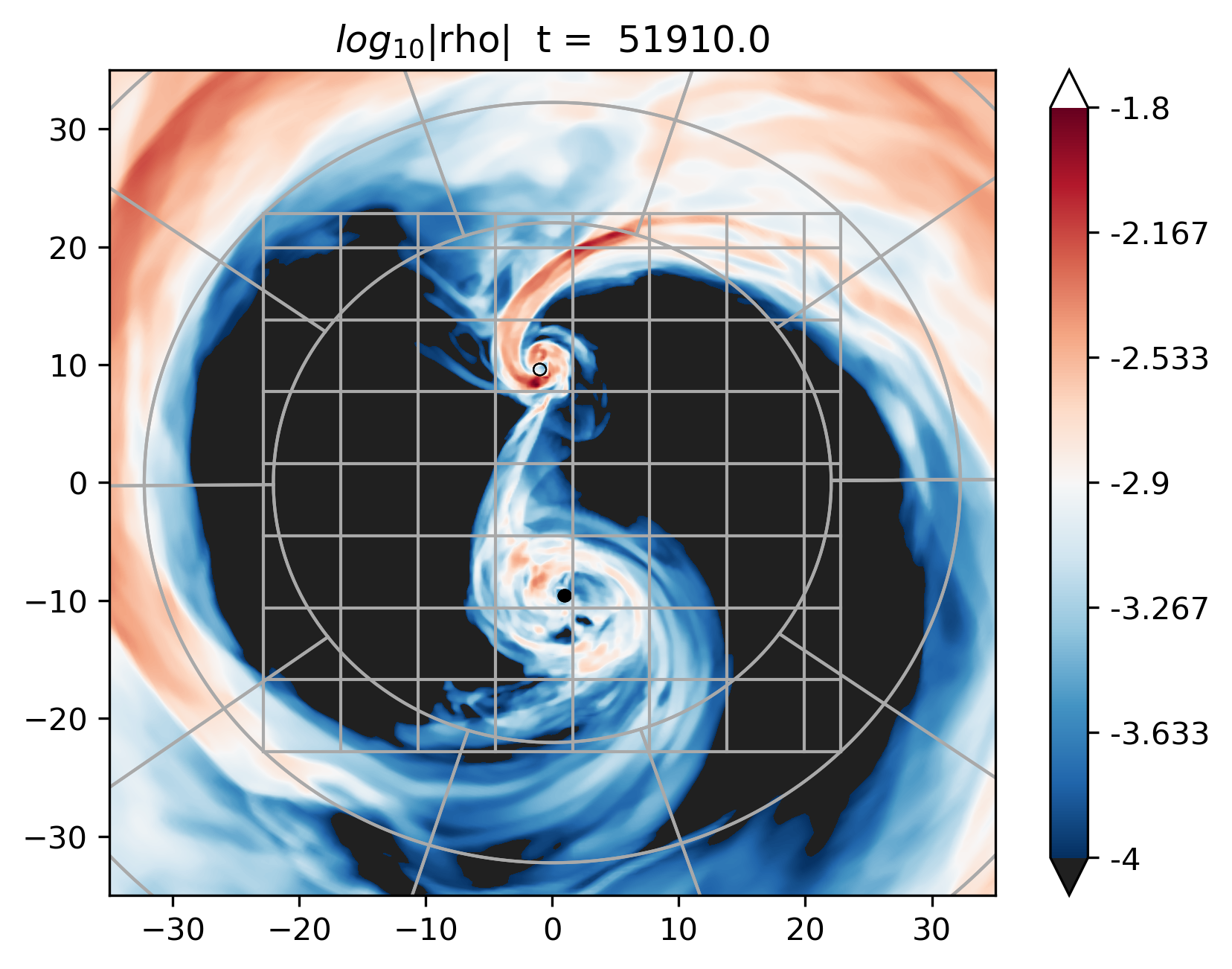}}
		\caption{An equatorial slice showing density (on a logarithmic color-scale) in the orbital plane as given by the moderate resolution simulation, PM.IN20s. {\bf [Left]} panel shows most of the radial extent and highlights the azimuthal overdensity, `lump' structure. The gray grids represent every 50th cell-boundary in the cartesian inner and polar outer patch. Where they overlap, the outer patch computes the system evolution. {\bf [Right]} is a zoom-in on the central region, the gray grid representing every 40th cell-boundary. BH1 and BH2 in both panels are identified with filled and open circles respectively. This snapshot taken just before $t\sim52000\mbox{M}$ is chosen so that BH1 is in the disk dominated state and BH2 accretion is stream dominated as it closely approaches the CBD lump. A movie of the simulation can be found at: https://youtu.be/q6bMg9CV0TA}
		\label{fig:equatorgridrho}
\end{figure*}

The older runs to whose data we compare our new results were conducted in the customary fashion, in which a single program evolved the entire problem volume.  However, our new runs use a multipatch method, in which the problem volume is divided into a number of ``patches" so that, taken together, the patches cover the entire problem volume.  Figure \ref{fig:equatorgridrho} shows the decomposition of our domain into two patches for our fiducial run.

Here we give only a succinct summary of the organization of our multipatch method.  Its basic structure is introduced in \cite{Shiokawa2018}. For this and forthcoming work, \pmhd demonstrates significant extensions and improvments leading to increased efficiency and applicability. Of neessity for binary accretion in 3D, \pmhd supports magnetic field evolution in ideal MHD. The primary changes in development of \pmhd are briefly summarized in the Appendix and will be given a more thorough description, including that of extensive testing, in a forthcoming paper (Avara et.~al. in preparation). 

In the combined \pwmhd+\harmd environment (from now on simply referred to as \pwmhd\ in this paper), there is an independent multiple program multiple data (MPMD) executable assigned to each patch, and these may therefore employ \harmd differently from one another. Each patch has its own spatial coordinate system and grid, but they must all share the same time coordinate.  For any cell covered by more than one patch, the program responsible for its update is determined by a pre-determined hierarchy.  The patches may or may not move relative to one another.  In order to reach a consistent solution over the entire problem volume, at each time substep, boundary condition data are exchanged across shared patch boundaries.  When patch A requires boundary data at a set of locations governed by patch B, the values of the primitive fluid variables in patch B are interpolated\footnote{A significant reduction in interpolation error has been achieved in the upgraded \pmhd by transforming vector quantities into the receiving patch coordinate system before interpolation, thus avoiding non-diagonal error multipliers. We highly recommend this for any interpolation of vector or tensor quantities in non-smooth media.} to the needed positions.

For the multipatch simulations presented here, we need only two patches.  One, using spherical coordinates, covers the region of the CBD, mimicking single-patch runs with the binary excised.  The other, using Cartesian coordinates, covers the binary itself, including the center-of-mass region (see Fig.~\ref{fig:equatorgridrho}).  It therefore eliminates the coordinate singularities afflicting any spherical coordinate representation. Where the two patches overlap, the spherical coordinate patch is chosen to compute the solution so that the inter-patch boundary minimizes boundary crossings and azimuthal symmetry is achieved in resolution of the accretion flow.   There is no relative motion between the two patches.

In previous work by our group in which we studied accretion through minidisks \citep{Bowen2017,Bowen2018,Combi2021}, we instead employed a single ``warped" grid whose  symmetry changed smoothly from spherical at large distances to quasi-Cartesian very near the two black holes.  Inherited from the spherical symmetry far from the center-of-mass, coordinate singularities remained along the polar axes and at the origin, necessitating radial and poloidal grid excisions.  This warped grid also demanded cells substantially smaller than the gradients in the variables required, and these excessively small cells led to very short time-steps.  By eliminating these constraints, our multipatch scheme both diminishes the computational cost by a factor $\gtrsim 30$ and provides the ability to capture gas flows through the center-of-mass region. This speed-up factor is estimated by comparing the per-orbit computational expense in CPU-hours of the single-mesh approach of \cite{Bowen2018,Combi2021} to that found in the \pwmhd\ run; this is a fair comparison because the new run has comparable or better numerical resolution.

Treating magnetic fields in such an approach creates a special difficulty in that the magnetic field is subject to an independent constraint: zero-divergence everywhere.  The FluxCT algorithm ensures that a field that begins with zero divergence everywhere in a problem volume interior acquires non-zero divergence only in its ghost cells; it is therefore a powerful tool for enforcing this constraint in single patch simulations.  However, the approximations inherent in any interpolation create divergence during the inter-patch boundary data exchange.  Put another way, connecting adjacent cells across a patch boundary inevitably leads to random errors that violate conservation of magnetic flux across that boundary.  Heuristically, one may think of the time-development of these errors as a random walk process that leads to steadily growing flux-conservation errors at the boundary, or incompletely closed field loops.  Although the FluxCT algorithm acting within the patches fixes the associated magnetic monopoles to the patch boundaries, their presence can cause inaccuracies in the evolution to bleed into the patches' interiors by generating erroneous magnetic forces on the fluid.

To solve this problem, we invented a routine to suppress the growth of magnetic divergence at patch boundaries by inserting into the FluxCT algorithm a damping term proportional to the local magnetic divergence.   This routine, along with another one that reduces magnetic divergence related to interpolation error arising when the patches move relative to one another, is described in the Appendix and, in greater detail, in the forthcoming methodology paper (Avara et al. (2023)).

\subsection{Details of the grid and boundary conditions}

Figure~\ref{fig:equatorgridrho} shows the midplane of the two stationary patches we use for this work, the inner Cartesian and the outer spherical-polar.
Where the grids overlap, MHD evolution is evaluated on the outer, spherical patch.

The actual code coordinates in all our simulations are called ``modified Kerr-Schild" (MKS). That is, they adopt the Kerr-Schild description of a Kerr spacetime, but the spatial coordinates (labeled $x^{(i)}$ for $i=1,2,3$) are not necessarily linearly proportional to either ordinary Cartesian or spherical coordinates.  In all cases, the MKS spatial coordinates internal to the code are discretized uniformly to minimize computational expense.

In the outer patch the grid is identical in shape to that used in \cite{Noble12}: logarithmic in the radial direction (i.e., $r \propto \exp[x^{(1)}]$ so that $\Delta r/r$ is constant) and uniform in azimuthal angle, but the polar angle cells are compressed near the midplane and stretched near the polar axis:

\begin{equation}
    \begin{split}
    \theta(x^{3}) = {\pi \over 2}\big[1 &+ (1 - \xi)(2x^{(3)} - 1)  \\
     &+ (\xi - 2\theta_c/\pi)(2x^{3} - 1)^n \big] .
     \end{split}
\end{equation}

Here the parameters $n$, $\xi$, and $\theta_c$ are, respectively, 9, 0.87, and 0.2.  The parameter $\theta_c$ defines the opening angle of the cut-out around the polar angle.  There are 260 (reduced from 300 in the source snapshot) radial cells (spanning the range from $22M$ to $260M$), 160 polar angle cells (154 for the high res run), and 400 azimuthal angle cells in the outer patch in all of the simulations reported here. For further details of the grid and physical setup of RunSE see \cite{Noble12}. 

The grid in the inner Cartesian patch is uniform in the orbital plane and occupies a range of $45.8M$ in both the $x$ and $y$ coordinates centered on the origin, which coincides with the binary center of mass (COM).  The cells in the $z$ direction span a distance of $200M$ centered on the orbital plane, but squeezed toward the plane in order to match the resolution in the outer patch, with roughly constant vertical cell aspect ratio of $\Delta\theta_{\mbox{outer}}/\Delta\theta_{\mbox{inner}}\sim4/5$ along that interface:
\begin{equation}
z(x^{(3)}) = 100\left[(1-\xi)(2x^{(3)} - 1) + \xi(2x^{(3)} -1)^n\right],
\end{equation}
where $\xi=0.96$ and $n=9$. This concentrates most resolution into the central cavity region, but leaves some room for structures extending farther from the midplane than the hydrostatic scale height associated with our cooling function.  As we report below, such structures appear.

The number of cells in each Cartesian patch coordinate varied between our different simulations.  In our fiducial resolution run, PM.IN20s, the number of cells in $x \times y \times z$ was $300 \times 300 \times 200$, but in our high resolution run, PM.IN20sHR, there were $600 \times 600 \times 300$ cells (see Table~\ref{tab:runs} for the full list of runs whose data are used in this paper).   Both versions led to sufficient resolution in the minidisks, as measured by community standards for cells-per-scale-height and quantitative similarity of the hydrodynamic evolution of PM.IN20s with PM.IN20sHR.  In PM.IN20s, the number of cells per vertical scale height in the initial state of the minidisks ranged from $\simeq 13$ near the ISCO to $\simeq 38$ at their outer edges.

Boundary conditions on the physical boundaries are pure outflow.   However, for the final third or so of run PM.IN20s, enough of the atmosphere of the central Cartesian patch has accreted that inflow from the upper and lower boundaries can become significant, bringing unphysical magnetic field into the domain.  To quell this, in the last portion of the run the pressure and density in each successively outer ghost zone was diminished by a factor of 0.8 relative to the next cell inward.  This device helps enforce outflow for the entire evolution. 

\subsection{Initial conditions}

The initial state of the CBD in all our runs was taken directly from RunSE in Noble12; the specific time whose data were copied is shown in Table~\ref{tab:runs}.  As is typical for CBDs, the disk's mass is crudely axisymmetric; its surface density rises sharply with radius near $r \simeq 2a$, reaches a maximum at $r \simeq 2.5a$, and declines gradually at larger radii.  These data were, however, somewhat modified for our present purpose. All cells on the outer patch with $r < 1.1a$ were removed so that the circulation of material in the Hill sphere of each black hole would take place entirely on the central patch and therefore be treated with approximately uniform resolution throughout its orbital motion. This adjustment also reduces any possible impact of interpolation error repeatedly affecting rotational flow in the minidisks, although as we will find, for this separation, there is little rotational symmetry. 

The initial magnetic field as also slightly modified in the innermost region. The starting snapshot of RunSE has magnetic flux threading the inner radial and axial boundaries, but the central patch starts with zero magnetic flux since we don't know {\it a priori} its structure. This discontinuity in the field would introduce a magnetic divergence into the domain, coincident with the interpatch boundary. To remove this, we first zeroed out the magnetic field in the radial range $1.1a \leq r < 1.21a$, creating an initially zero-field buffer between magnetized flow and the interpatch boundary. This way our divergence cleaning algorithm did not need to span the inter-patch boundary. We also set the magnetic field to zero in cells for which $|\theta - \pi/2| > 0.16\pi$. This latitudinal width was chosen to be small enough so as not to disturb the more strongly magnetized material near the surface of the accretion disk, but large enough so that subsequent divergence cleaning does not re-introduce a significant field crossing the boundary. 
After making these changes, we used the magnetic divergence removal routine of \cite{Bowen2018}, a successive over-relation variant of the Gauss-Seidel algorithm, to minimize divergence along the new field discontinuity.  This process leads to negligible divergence in both the outer patch and at the inter-patch boundary, and does not significantly alter the magnetic evolution, especially further out in the CBD.
The only other change made to the MHD quantities in the outer patch was a renormalization of the magnetic field by the factor $\sqrt{g_\mathrm{old}/g_\mathrm{new}}$, where $g_\mathrm{old}$ ($g_\mathrm{new}$) is the determinant of the old (new) metric, and the updated metric includes the inner zone (perturbed Schwarzschild metric) spacetime contribution not needed in the cavity-excised runs of Noble12. The renormalization creates canceling contributions to the FluxCT representation of the magnetic divergence, so no additional divergence cleaning is necessary.

The region closer to the center-of-mass than the innermost location in the grid of Noble12 was filled with low density material.  Having observed the rapid filling and draining of the minidisks in \cite{Bowen2019}, we were confident that this choice would lead to reaching the minidisk mass accretion cycle several orbits sooner than by initially endowing the minidisks with substantial gas \citep{Combi2021,Gold2014a}.  

The low density ``atmosphere" has a density of $1 \times 10^{-8}$ in code units everywhere inside $r=40M$, much smaller than that of most dynamically relevant material, for which $\rho\sim10^{-4}-10^{-1}$.  Outside $40M$ the ``atmosphere" density diminishes $\propto (r/40M)^{-1.5}$.  Similarly, the atmosphere's internal energy density is $1 \times 10^{-10}$ for $r < 40M$, but decreases $\propto (r/40M)^{-2.5}$ at larger radii.   These values were reduced to $8.8 \times 10^{-12}$ and $1.1 \times 10^{-14}$, respectively, just before the end of the first orbit, when the boundary conditions on the top and bottom surfaces of the Cartesian patch were altered to reduce inflow across these boundaries. The lower density and pressure also reduce inflow at those boundaries.  These atmospheric profiles are the same as the floors enforced on the primitives during evolution to avoid U(P) inversion failure.

\begin{table*}[t]
	\centering                                                                                                                                                     
\begin{tabular}{
ccccccccr}
	\hline
	\noalign{\vskip 0.1cm}
	\textbf{Run Name}    & Code          & Duration [$10^3$ M] & Sep. & $t_{\mbox{shrink}} [M]$ &  Cooling   &  Resolution &  Metric & Source \\
	\noalign{\vskip 0.1cm}
	\hline
	\hline
	\noalign{\vskip 0.1cm}
	RunSE        		 & \harm 		 & 0 - 75      & Fixed		& Never    &  Bound  & 300x160x400 &  NZ    & Noble12  \\
	RunIN         		 & \harm		 & 40 - 54     & Evolves		& 40000 &  Bound  & 300x160x400 &  NZ     & Noble12  \\
	PM.IN20s     		 & \pwmhd  & 50 - 64     & Evolves		& 50000 &  Bound  & 300x300x200 &  NZ+IZ	& new  \\
	PM.IN20sHR    		 & \pwmhd  & 50 - 55     & Evolves		& 50000 &  Bound  & 600x600x300 &  NZ+IZ	& new  \\
	PM.IN20s-CuB  		 & \pwmhd  & 55.1 - 56.0 & Evolves		& 50000 &  All    & 300x300x200 &  NZ+IZ	& new  
\end{tabular} 
	\caption{This table provides a summary of the simulations included in the analysis and discussion of this work.  Note that runs with names of the form ``PM.IN$\ldots$" used $\pwmhd$ combined with \harm, and covered both the minidisks and the CBD; the others covered only the CBD and used only \harm. Columns include the following descriptors: run name; the duration of the run; whether the separation of the binary evolves or is fixed at 20M; the time in RunSE from which its initial data were taken; whether only bound material or all material is cooled; cell counts for $r \times \theta \times \phi$ for CBD runs or $x \times y \times z$ for the Cartesian patch in \pwmhd runs; the metric used, with naming convention as in \cite{Farris2014}.}
	\label{tab:runs}
\end{table*}

\section{Global accretion behavior: Time-dependence, Disk states, and 3D effects}
\label{sec:hydroresults}

In this section, we focus on how matter flows from the CBD to the minidisks and then to the black holes as the binary separation shrinks from $20M$ to $\simeq 9M$.   Previous efforts studying this problem \citep{Bowen2018,Bowen2019,Combi2021} were limited by relatively short duration: 3, 12.5, or 15 (12.5 for spinning BHs) binary orbits for these three papers, respectively.  During this relatively short time, the binary orbit evolved at most modestly.  Because the work we present here covers 36 orbits, we are able to explore minidisk dynamics over more than a factor of 2 contrast in binary separation, covering the last 98\% of the remaining time before merger from this initial separation.  Earlier efforts were also hampered by the presence of grid excision enclosing the center-of-mass region, whereas this region is fully included in our new work.   Both improvements in treatment are made possible using \pwmhd. 

For orientation we return to Fig.~\ref{fig:equatorgridrho}, comprising two images of the gas density in the orbital plane at a time shortly after the end of initial transients in our fiducial simulation.  As can be seen in the zoomed-in figure on the right, there is a sharp truncation in density on an irregular surface at roughly twice the binary separation.  Inside this edge, the density drops by an order of magnitude or more, while in the factor of $\simeq 2$ in radius outside the break there are large-amplitude spiral waves imprinted on the CBD.  Moreover, this disk is both eccentric and, due to the lump mentioned earlier, strongly non-axisymmetric.  Two stream-like structures cross the low-density gap inside the inner edge of the CBD.  One of them carries much more mass than the other and is much less diffuse; both properties arise from its proximity to the lump, which can be seen on larger scales in the left panel of Fig.~\ref{fig:equatorgridrho}. 
Later in this section, we will describe how the minidisks cycle between two distinct accretion states.
The minidisk being fed by the more massive stream is, at the moment of this snapshot, near a minimum in total mass and a peak in accretion rate.  The state of each minidisk flips on a timescale $\sim 0.7\times$ the binary orbital period.  The two minidisks interact as mass ``sloshes" \citep{Bowen2017} from one to the other; which way this mass travels also follows the accretion state cycle.

\subsection{Accretion rate}
\label{sec:accretionrate}

\begin{figure*}
	\includegraphics[trim={0cm 0cm .5cm 0cm},clip,width=2.06\columnwidth]{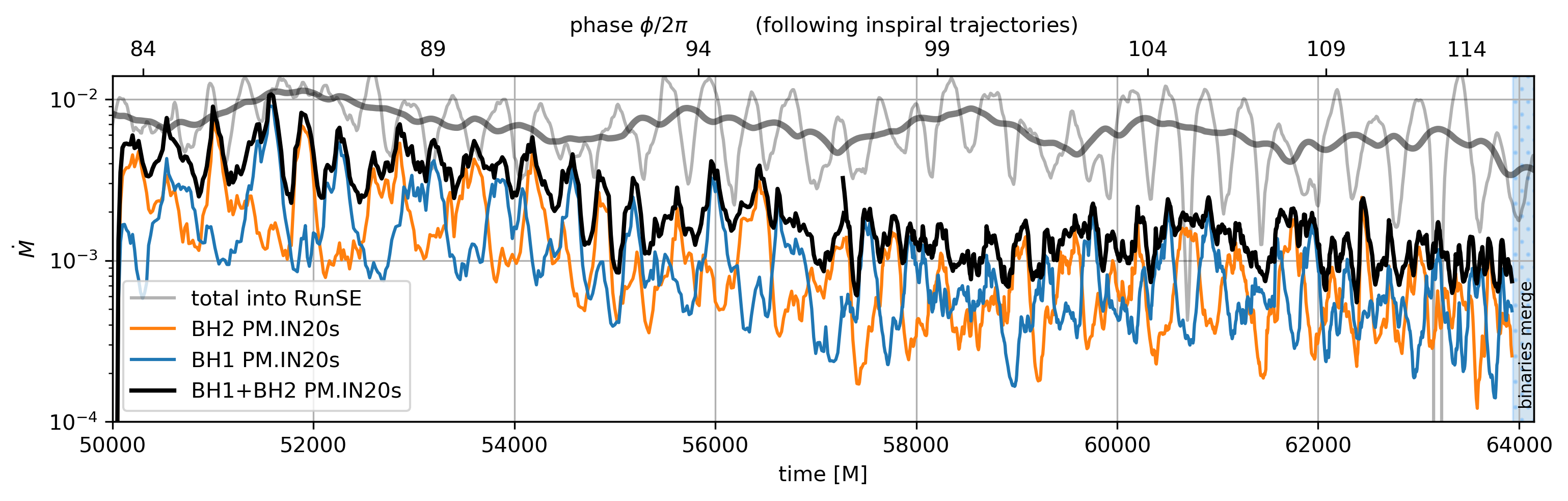}
	\caption{
		The accretion rates onto BH1 (blue) and BH2 (orange) measured just outside their respective horizons in the fiducial simulation PM.IN20s; their sum is shown by the thick black line.  For comparison, the total accretion rate \dotm\ at the inner grid radial boundary, the `cut-out' of RunSE in Noble12 at $r=0.75a=15M$, is shown by the light-gray curve; a smoothed version is shown by the thick-gray curve.  Time in units of $M$ is given along the bottom edge, and time in units of binary orbits is along the top edge (the separation in time can be found along the top edge of Fig \ref{fig:coolfuncvstselect}, which plots effective light curves from energy dissipation). The last (blue-hatched) shaded region shows the full remaining time before the merger. Numerical evolution of Einstein's equations would be needed to extend the simulation here. }
	\label{fig:MdotandRunSE}
\end{figure*}

\begin{figure}
	\includegraphics[width=\columnwidth]{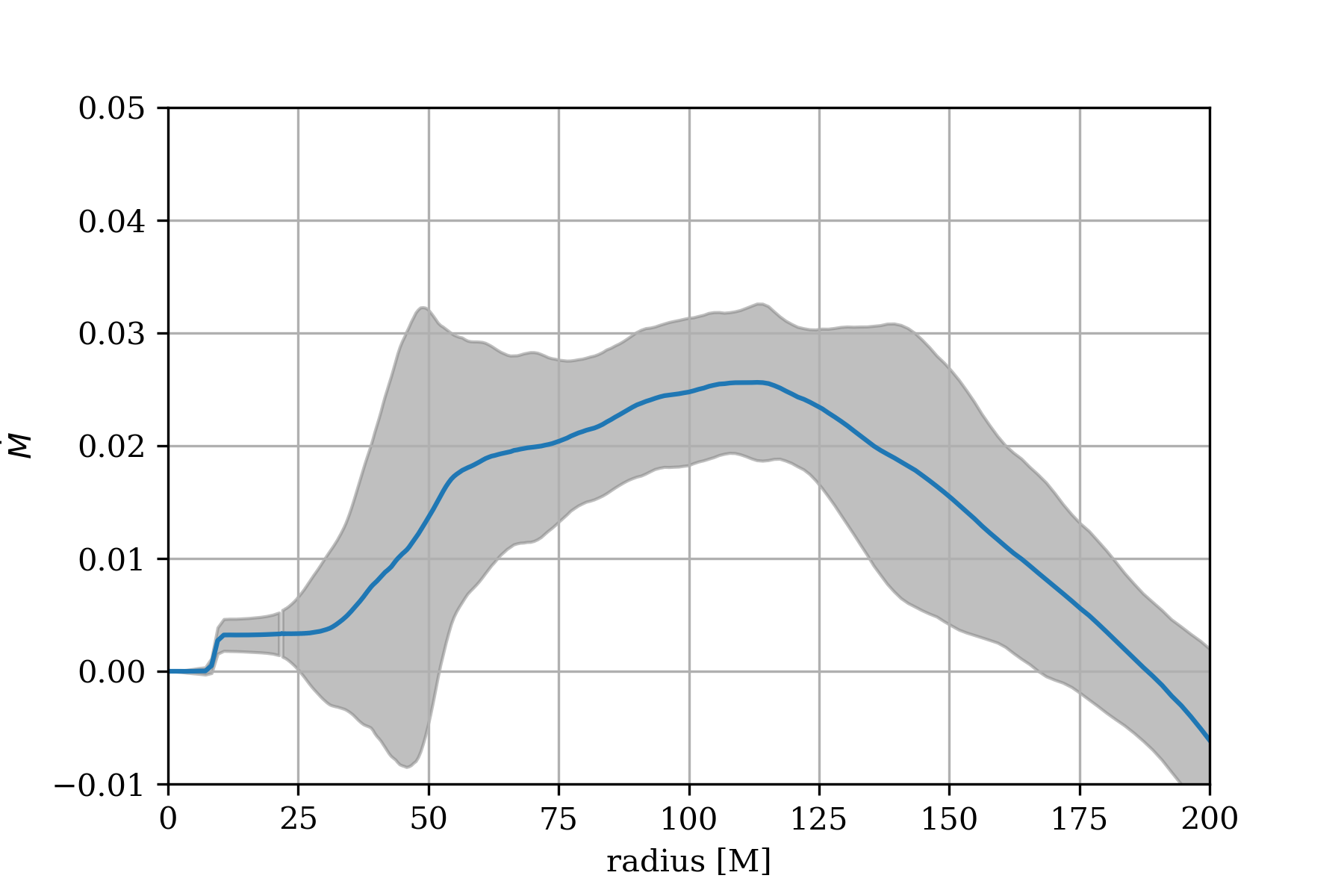}
	\caption{$\dot M$ as a function of radius (in units of $M$), averaged over $t=[50000,60000]$M (blue line). This epoch corresponds to the first $\sim 20$ binary orbits of the simulation.  By averaging over an integer number, 4 complete lump orbits, we minimize the impact on the time average of lump-driven modulation. We do not include the last $\sim4000\mbox{M}$ in the average since the radial scale changes much more significantly over that period of inspiral. The gray-swath indicates the $1 \sigma$ width of instantaneous fluctuations around this mean.
	}
	\label{fig:mdotvsrfull}
\end{figure}

The most fundamental quantity one can measure in an accretion flow is the rate at which mass moves toward the central gravitating object(s).  We summarize our results in this respect by two measures: the accretion rates onto the two black holes as functions of time (Fig. \ref{fig:MdotandRunSE}), and the time-averaged accretion rate as a function of distance from the system center-of-mass in Fig. \ref{fig:mdotvsrfull}.
Note that the units for accretion rate are code-units, not physical. We use these units to facilitate comparison to prior studies with similar CBD conditions, and the units for its mass are arbitrary because the absolute amount of gas mass does not influence the dynamics. To make predictions about specific cases in physical units, all that is necessary is to identify a particular value of $\dot M$ in code-units with a physical rate while also choosing the total mass of the binary \citep{Schnittman+2006}. \footnote{In practice, the choice to artificially cool the disk to achieve a geometric target thickness is a sort of implicit choice for how radiation will effect energy transport and pressure balance, but the actual effects of radiation on accretion flows remain so inadequately understood that nearly all normalizations of physical units are equally motivated.}

At the very beginning of PM.IN20s, the accretion rate measured at the inner edge of the CBD (r=1.1a(t=50000M)=22M) is in excellent agreement with its pre-equilibration simulation RunSE, in which the binary separation did not evolve.
The mass flux through the inner edge of the CBD in PM.IN20s then drops compared to RunSE during the initial transient period by a few tens of percent, possibly due to decreased stress in the region where magnetic field is initially removed. Although the accretion rate in RunSE declined by a factor $\sim 2$ over the subsequent $\sim 14000M$ in parallel time, the rate measured at r=22M in PM.IN20s declined by a factor $\sim 3-4$ in its first $\sim 7000M$, but then stayed nearly constant for the remainder of the simulation---except for a brief upward fluctuation by a factor $\sim 2$ shortly before the simulation's end.

In RunSE, there was, of course no way to measure the accretion rate onto the black holes. Here, we find that the accretion rate onto the black 
holes matches that at r=22M very closely through most of the evolution, and even better matches the value at the shrinking inner edge of the CBD. The combined accretion rate onto the black holes measured near the horizons drops by a factor of $\sim 4$ and then remains steady at this value until there is a slight drop in the final $\sim 1000M$ prior to merger.

The most natural interpretation of the greater drop in accretion rate found in the new simulation is that it is due to the shrinkage of the binary.  As the binary separation decreases, its quadrupole moment diminishes, permitting stable quasi-circular orbits to exist at smaller radii in the CBD. To the degree the inner edge of the CBD moves inward more slowly than the binary compresses, this effect reduces the accretion rate from its inner edge.  Unlike early estimates \citep{MP05}, however, this is not a ``knife-edge" effect; hence, the reduction in accretion rate is only by a factor of order unity.

As previously seen in numerous simulations (e.g., that of \citet{Noble12}), the long-term trend in accretion rate in PM.IN20s is modulated periodically on a timescale $\sim 5$ binary orbital periods.  However, the amplitude of this modulation decreases substantially, becoming imperceptible roughly halfway through the simulation.

\begin{figure}[]
	\includegraphics[width=\columnwidth]{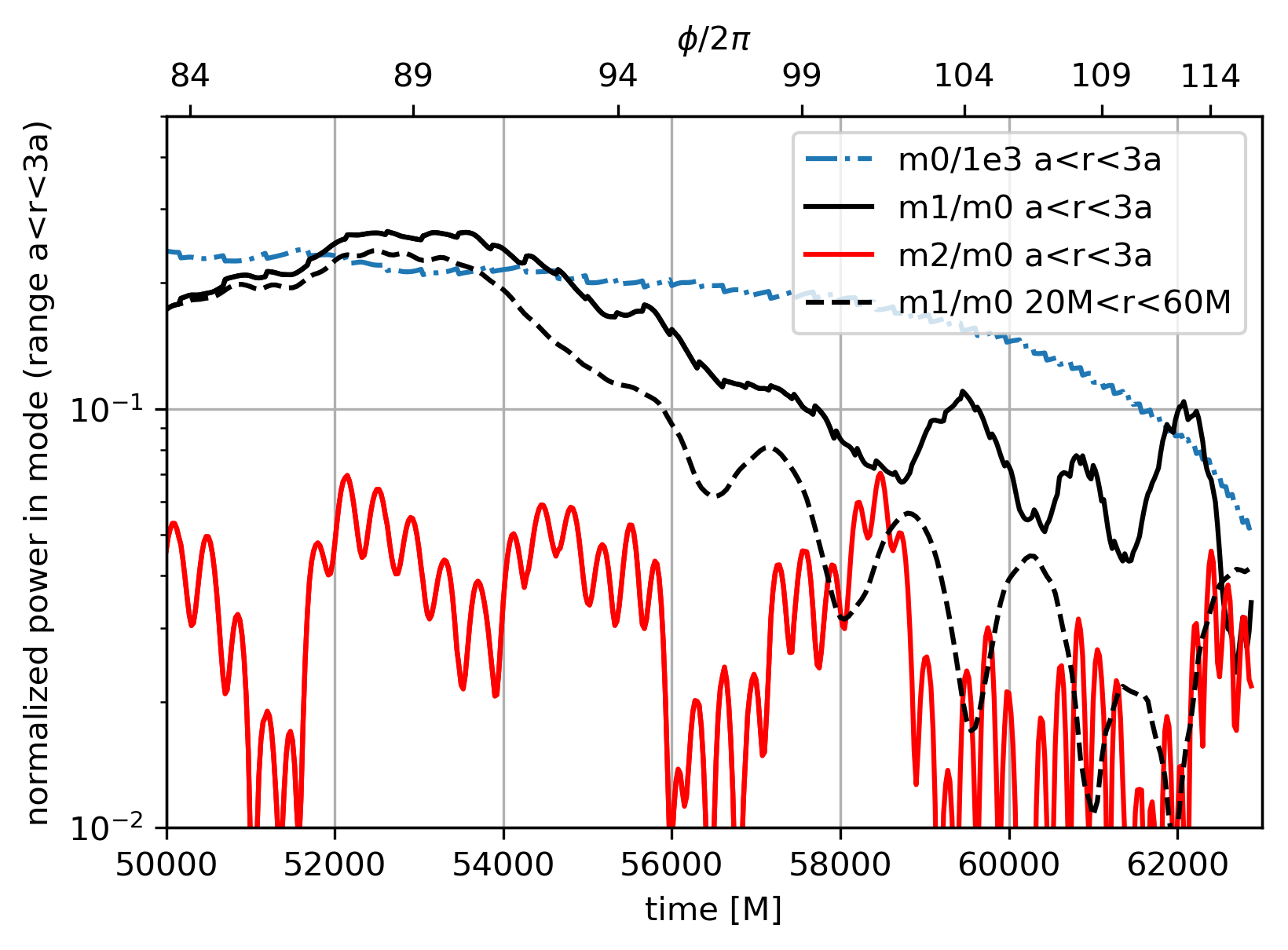}
	\caption{The first three azimuthal modes of rest mass of the CBD inner edge (measured between radii $22.0M$ and 3a ) are plotted in blue-dash-dot, black-solid, and red-solid respectively. The 1st and second azimuthal modes are normalized by $A_{m=0}$, the total integrated mass over that radial range. The black-dashed line is the power in $m=1$ radially integrated over a fix lab-frame range, whereas the others are fixed with respect to separation a(t). } 
	\label{fig:CBDmodesvst}
\end{figure}

The transition seen to occur around $t=56000M$ (orbital phase of $\sim95\times2\pi$)---from declining to nearly constant $\dot M$\ and the disappearance of the $\simeq 5$ binary orbit modulation---is likely due to orbital evolution undermining the mechanisms responsible for lump and eccentricity reinforcement \citep{Shi12}.  As shown in Figure~\ref{fig:CBDmodesvst}, shortly before this time the amplitude of $m=1$ azimuthal modulation of the CBD surface density begins to decline sharply.  In other words, the lump becomes progressively less prominent.  During this period, CBD mass spreads into the decoupling gap (the gap between the CBD truncation edge at earlier times and its current radial location, now filled with stable quasi-circular orbits) and stretches azimuthally due to orbital shear.  As is also shown in Figure~\ref{fig:CBDmodesvst}, the radial component of the spreading and partial decoupling of the binary from the CBD also leads to a decrease in surface density in the radial range $a <r < 3a$, particularly after $t \simeq 60000$.

As shown by \cite{Noble21}, when the binary mass-ratio $M_2/M_1$ is less than a few tenths, the weaker quadrupole of the binary is no longer able to support lump formation for our choice of disk thermal state (i.e., scale height). Our results indicate that even when the binary is equal-mass, a weaker quadrupole at a given fixed radius in the CBD due to rapid inspiral also undermines lump reinforcement.

The time-averaged accretion rate as a function of radius from the binary center-of-mass is shown in Figure~\ref{fig:mdotvsrfull}.  The relatively flat profile for $\dot M$, albeit with a large variance, between $r=50M$ and $r=125M$ reflects the inflow quasi-equilibrium attained in the CBD by using the pre-evolved simulation from Noble12. The particularly large variance near the CBD inner edge ($r \approx 40 - 50M$), even showing net negative mass flux at times, is due to the large-amplitude spiral density waves induced by the time-varying binary quadrupolar moment, and the radial oscillation of the lump on its eccentric orbit (see Fig.~\ref{fig:equatorgridrho}).  The smooth decline in accretion rate from $\simeq 50M$ to $\simeq 30M$ implies the progressive filling-in of this region as the inner edge of the CBD tries to keep up with the shrinking binary separation. Analysis in comparison with RunSE reveals that out of the full factor of 10 difference between equilibrated radii of the CBD and the minidisks, a factor of about 2-3 can be attributed to the lump still growing in the early portion of the simulation. 

In evaluating the time-averaged accretion rate as a function of radius, it must be borne in mind that during the averaging period the accretion rate from the CBD's inner edge and through the minidisks diminishes by a factor of $\sim 4$ (see Fig.~\ref{fig:MdotandRunSE}).  At $t=50000M$, the accretion rate into the gap is smaller than the accretion rate at $r \sim 100 \pm 50 M$ by about a factor of 3-4 (0.006-0.008 vs. $\sim$0.024); the ratio grows by a similar factor by the end of the simulation.  During this comparatively short time, the disk at larger radii cannot re-equilibrate. Finally, this measure of $\dot M$ goes to zero at small radii between the BHs because, of course, this is the net flow across a closed surface that does not include a gravitating point mass. 

\subsection{Accretion cycles between states}
\label{sec:accretionstates}

It has been previously noted \citep{Bowen2018,Bowen2019,Combi2021} that the time fluid elements spend in a minidisk is comparable to or shorter than a binary orbital period when the binary separation is $\sim 20M$ and the BHs are non-spinning (also noted by \cite{Gold2014a}, but observed at a separation $\sim 10M$). Because the lump-fed accretion stream alternates between feeding the two minidisks, there is a large ratio between the mass of a minidisk immediately after it has completed receiving its supply and its mass shortly before a new delivery begins. The complete cycle between these states is largely independent of the driving timescale, which evolves from $\approx 1.4$ binary orbits at early times to $\approx 1$ orbit about halfway through the simulation.  Earlier, the modulation is driven by the lump's orbit; later, as the lump amplitude decays, it is instead driven by the eccentricity of gas orbits at the CBD's inner edge.
First, {let us} present a deeper analysis of how a minidisk's structure changes as its mass varies through a full accretion cycle using purely hydrodynamic behavior, and in \S\ref{sec:magstates} we will dive further into the cyclical magnetic component. 

\begin{figure}[]
	\begin{centering}
		\subfloat[Disk dominated: $t=56025M$]{
			\includegraphics[width=1\columnwidth]{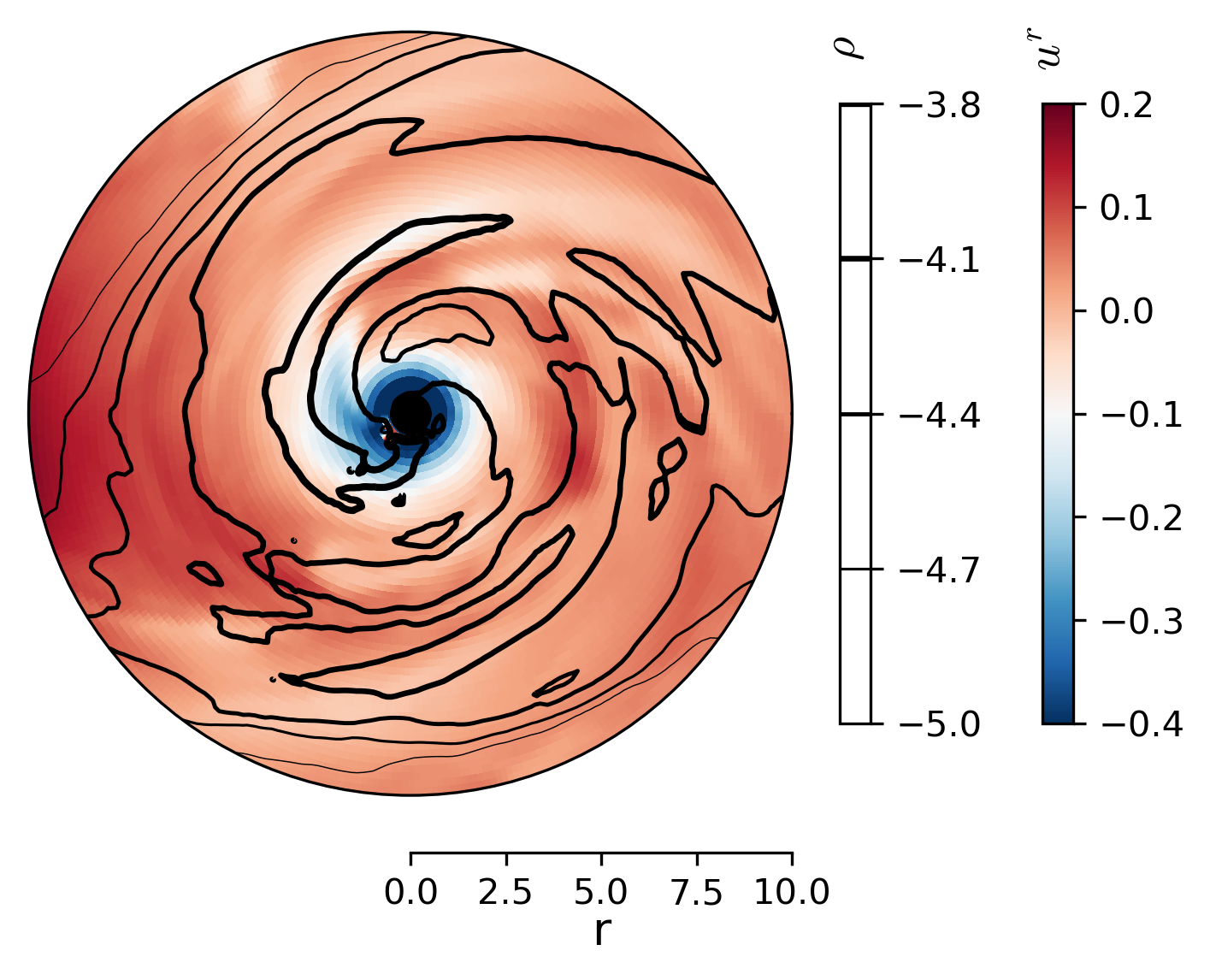}}\\
		\subfloat[Stream dominated: $t=55600M$]{
			\includegraphics[width=1\columnwidth]{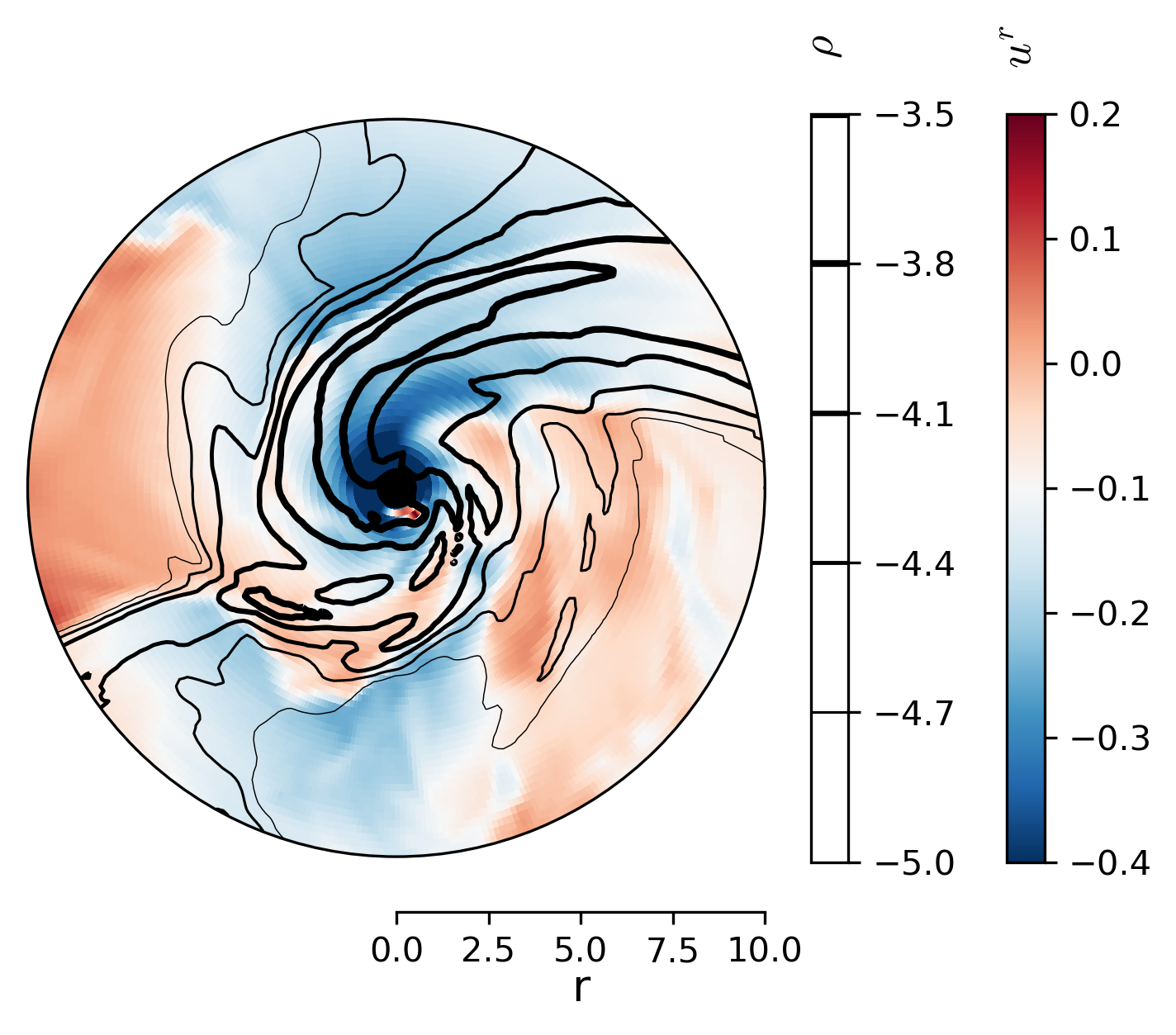}}
		\caption{Slices through the equatorial plane of {BH2} minidisk at {\bf(a)} $t=56025$M , and {\bf(b)} $t=55600$M. Contours are $\text{log}_\text{10}(\rho) $ where $\rho$ has been vertically averaged over the central 4 scale heights {spanning} the equatorial plane. Note that the density in the stream in (b) peaks at a value higher than in (a), {so an additional density contour has been added at the upper end of the range.} The color-scale, centered on -0.1, shows $u^{r}$. }
		\label{fig:strmvsdisk}
	\end{centering}
\end{figure}

{  {Figure~\ref{fig:strmvsdisk} shows how different the maximum and minimum mass states are.  In the upper panel the BH2 minidisk is ``disk-dominated": its mass is large and the accretion rate onto it is somewhat less than the maximum (BH identity as indicated in Fig.\ref{fig:equatorgridrho}).
In the bottom panel (``stream-dominated") this same minidisk contains little mass, but the rate of accretion onto the black hole is near the maximum rate.}} 
When the disk is full, accreting matter arriving at the disk is pushed onto roughly circular orbits; by contrast, when the disk has been depleted, arriving matter falls almost directly into the black hole.  

In the disk-dominated state, mass { following quasi-Keplerian orbits is} spread throughout the minidisk's Roche lobe,  although a one-armed spiral wave is evident.  By contrast, in the stream-dominated state, nearly all the minidisk's mass is in a curving stream that traverses only half a circuit around the black hole before plunging in, {thereby avoiding circularization through self-intersection}.  However, because the material is concentrated in a much smaller area, the maximum density in the stream-dominated state is a factor of $\simeq 3$ greater than in the disk-dominated state.
The two states also differ sharply in how rapidly the gas in a minidisk moves radially inward.  In the stream-dominated state, most of the material in the stream component has radial velocity $u^r < -0.1$.
The colorscale in Figure~\ref{fig:strmvsdisk} is therefore chosen to pass through white at this value, {distinguishing the stream and disk components.} In sharp contrast, during the disk-dominated state, nearly the entire minidisk has an inward radial speed with magnitude $\ll 0.1$.


\begin{figure}
  \begin{centering}
		\includegraphics[width=\columnwidth]{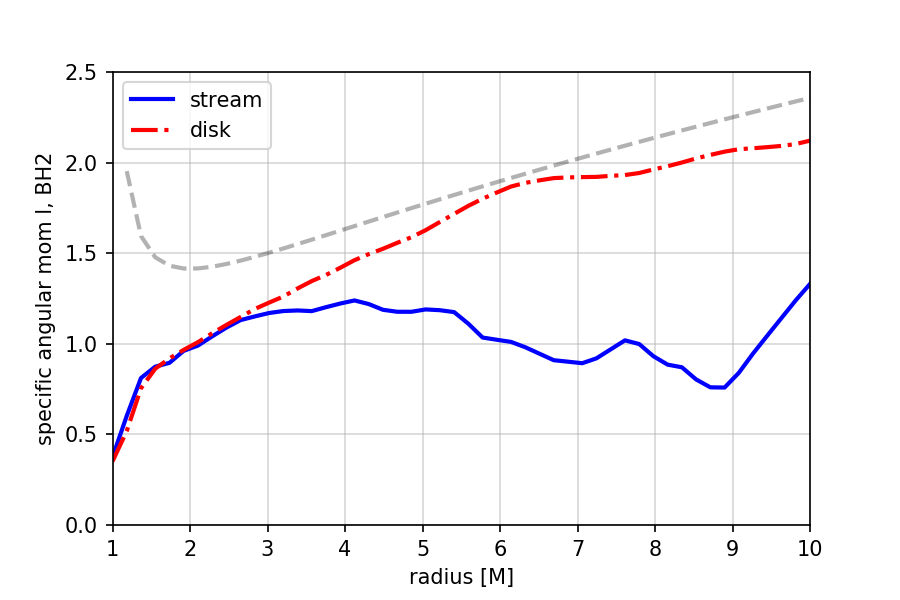}
		\caption{Minidisk specific angular momentum in boosted accelerating frame of BH1, vs. the radial coordinate centered on BH1, in units of the total binary mass M. Blue-solid line is the time average over the entire minidisk, but only including times when at least 70\% of the disk mass is streaming, as identified where $u^r\ll-0.1$, i.e. the minidisk is in the stream-dominated state. Red-dash-dot line is the analogous average taken over times when less than 30\% mass is streaming, thus times when accretion is disk dominated. Specific angular momentum of quasi-Keplerian circular orbits is gray-dashed. In these coordinates the ISCO occurs at $r=2.5M$, while the tidal truncation radius is at $r \simeq 8M$. } 
		\label{fig:kepvsr}
  \end{centering}
\end{figure}

The stark differences between these states stem from a combination of the large amplitude variation in mass-supply rate and the comparatively low specific angular momentum of the mass delivered to the minidisks.  The latter fact is demonstrated in Figure~\ref{fig:kepvsr}.  As also noted by \cite{Combi2021}, the specific angular momentum of the matter arriving at the minidisk---relative to the black hole it is about to orbit---is in general less than what is required for a circular orbit at a minidisk's tidal truncation radius.

{In fact, as shown by Figure~\ref{fig:kepvsr}}, the mean specific angular momentum of mass in the stream at the tidal truncation radius of the minidisk ($r \simeq 8M$ in the coordinates of the simulation before significant binary tightening) is already less than that of an ISCO orbit. Only a small portion of this material circularizes or comes in with high enough angular momentum to form a disk. Even this higher angular momentum matter quickly loses angular momentum to spiral waves and, in some cases, more stationary shock features (see next subsection).

Finally, as we will see in the subsection on tilts, the circularization of infalling material is further complicated by vertical structure and variability of the minidisks.

\subsection{Sloshing}
\label{sec:sloshing}

When discussing the global features of this binary accretion flow, we pointed out the mass exchange between the two minidisks shown clearly in Figure~\ref{fig:equatorgridrho}. In fact, that significant mass can be transferred from one minidisk to the other was already pointed out by \cite{Bowen2017}, even though the simulations of that paper contained a sizable cut-out in the path of this flow.  Our unbroken evolution of the center-of-mass region allows us to make a much more quantitative study of this ``sloshing" motion. 

\begin{figure*}[]
	\includegraphics[trim={0cm 0cm 1cm 0cm},width=2\columnwidth]{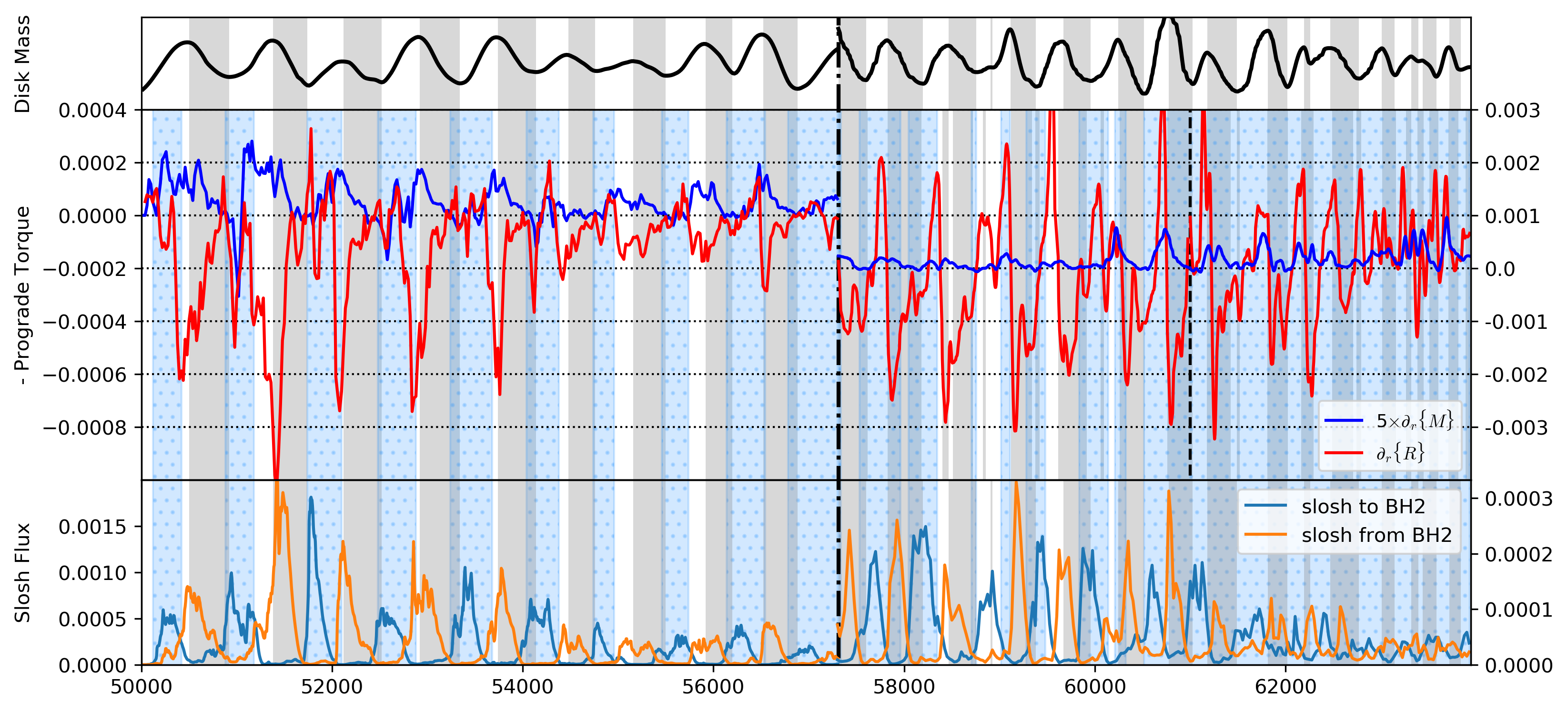}
	\caption{
	{[Top Panel] Variability of the BH2 minidisk's mass (arbitrary units), smoothed and de-trended to illustrate cyclical behavior (see footnote for details). Gray swaths across all 3 plots correspond to the times of decreasing BH2 mass. Blue-hatched swaths cover epochs where at least 25\% of the BH2 minidisk mass is located in the plunging streams, making the minidisk  `stream-dominated'. [Middle Panel]
 {  The shell-integrated torques produced by Maxwell and Reynolds stresses are $-\partial_r\{M\}$ and $-\partial_r\{R\}$; however, positive torques lead to outflow rather than accretion.  We therefore plot $\partial_r\{M\}$ and $\partial_r\{R\}$, radially averaged over the BH2 minidisk, with blue (Maxwell) and red (Reynolds) lines; for these quantities, positive values connote inflow.}
 The Maxwell torque magnitude is shown with $5\times$ its actual value to make its variations visible in comparison to the Reynolds values.
 At each time in the range $t=[50000,61000]M$, the plotted values indicate radial averages taken over $r_{\rm BH2}=[4,8]M$ (measured radially from BH2, in units of total binary mass $M$); after 61000M the radial average is over $r_{\rm BH2}=[4,6]M$ to account for the reduced minidisk size. For visual clarity, 
 the vertical scale for all curves to the right of the vertical dash-dot line is found on the right edge of the figure.
[Bottom Panel] Sloshing mass-transfer rate to BH2 from BH1 (blue line), and to BH1 from BH2 (orange line).}
    }
\label{fig:sloshmdotbh2}
\end{figure*}

Figure~\ref{fig:sloshmdotbh2}  (lowest panel) shows that, like the accretion rate, the sloshing rate proceeds by quasi-periodic pulses rather than a continuous flow.  The sloshing pulses are strongly correlated with the accretion rate and disk state, as shown by comparison with the accompanying panels including the trend in total disk mass\footnote{{Integrated mass across the minidisk is de-trended over long timescale variation of amplitude, and smoothed over individual  cycles as follows: The variation occurring slower than the orbital period is de-trended in both structure and variance using polynomial fits of 6 and 3 degrees respectively. Over sub-cycle periods this is followed by a smoothing Savitzky-Golay filter to make gradients monotonic across accretion cycle trends.} } and in relation to Figure~\ref{fig:MdotandRunSE}.  When a minidisk is transitioning from the stream-dominated to the disk-dominated phase, its extent grows because it is accumulating and circularizing mass.  More mass is therefore placed higher in the potential around the black hole as angular momentum is transferred into this material from closer to the BH.  The torque on the gas also does work, so that some of the mass reaches the orbital energy required to cross through the L1 point between the black holes.  For this reason, the main sloshing pulse leaving a minidisk tends to occur near the time it is disk-dominated and massive.

At this time, the other minidisk is just beginning to become stream-dominated, and is nearly depleted in mass.  When the sloshing pulse reaches it, the impact helps propel some of its remaining gas into that black hole, starting the next impulsive accretion episode.  {As a result, the  peak accretion reached rate at the horizon in each cycle depends on} the kinetic energy and timing of sloshing impact, as well as the quantity of material streaming inward in the subsequent stream-dominated state. {The relative timing of these two contributions to the high-acccretion phase of the cycle contributes to the complex variability across the peak, sometimes creating two close sub-peaks in \dotm.}

During the first $\simeq 11000M$ of our simulation, when the total accretion rate onto the binary is gradually decreasing and then reaches a plateau,
the mass transferred per pulse is $\sim 0.1 - 0.2 \times$ the mass accreted during the associated accretion pulse.  However, during the
last $\simeq 3000M$, although the accretion rate is roughly constant, there is visual evidence for less circularization {by} stream self-intersection,
and most streaming material simply accretes by plunging past the ISCO.
Throughout this phase, the ratio of the sloshing rate to the accretion rate is only $\sim 0.05$. In addition, unlike prior stages of evolution when the sloshing is nearly 100\% uni-directional at any given time, this final stage exhibits two-way exchange of material even when one dominates the sloshing flux. {This is indicated in Fig. \ref{fig:sloshmdotbh2}'s last panel, where neither sloshing flux line regularly returns to 0.  }

Total sloshing mass flux is not the only distinguishing measure of its impact on the minidisk states. The kinetic energy carried in the flow can be significant.  To estimate how much is available for dissipation through shocks, compression heating, turbulent dissipation, etc., we integrate the kinetic energy flux across the same surface as the sloshing mass flux.  In Fig.~\ref{fig:coolfuncvstselect} we plot the kinetic energy flux as an efficiency, normalized by the accretion rate onto the BH receiving the sloshing mass.   For nearly the entire run ($50000M < t < 62000M$), the mean kinetic energy flux is roughly half the photon luminosity of the recipient minidisk, but for brief times can even be greater.   These moments tend to occur at the beginning of radiatively bright phases of the minidisks, corresponding to peaks in the accretion rate and minidisk growth. The overall sloshing is correlated with the early minidisk dissipation in the radiative cooling, and in many cycles causes an obvious leading peak in the total synthetic lightcurve of the minidisk, followed by the primary stream-driven maximum. 

\begin{figure}[]
	\includegraphics[width=\columnwidth]{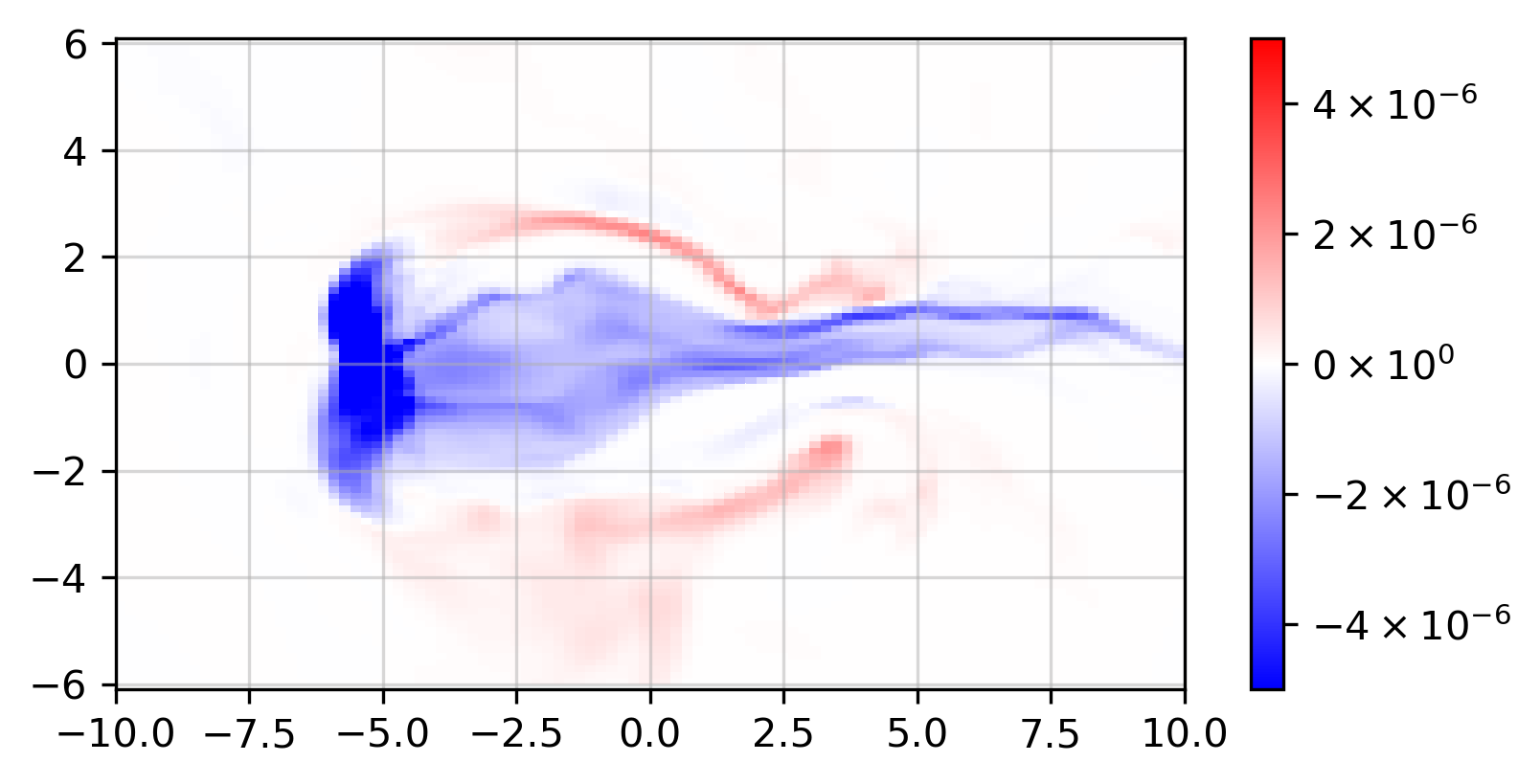}
	\caption{{Snapshot of mass flux at $t=52000M$.  {  The image plane is defined by the normal to both the binary orbital plane and the line between the two black holes,} the y and x axes respectively. Spatial units are [M]. {  Color density shows sloshing rest mass flux  $\rho_0 u^\perp$ passing through the image plane.}  Positive flux is toward BH1. To find $u^\perp$ we transform from r-$\theta$-$\phi$ coordinates to x-y-$\perp$. The extent of the plane was chosen so that it captures $\gtrsim95$\% of the total flux.}} 
	\label{fig:sloshprofile}
\end{figure}

\subsection{Summary of the accretion cycle}
\label{subsec:sumcyc}

With this quantitative view of sloshing in hand, we can complete the description of the minidisk accretion cycle. Figure~\ref{fig:sloshmdotbh2} presents a combined picture of the cycles driven in BH2, relating the oscillations of total disk mass to both sloshing and minidisk accretion torque {(detailed discussion of how torque is calculated is saved for \S\ref{sec:magnetic}, but for now it is sufficient to distinguish between the Reynolds $r-\phi$ stress $R$ at a point in the minidisk, its shell integrated value $\{R\}$, and the negative radial gradient of this value, $-\del_r\{R\}$, which {  gives the torque on material of that shell, i.e., when positive, the matter's angular momentum increases.}
{  In order to emphasize processes driving accretion, we therefore} plot $+\del_r\{R\}$ in Fig. \ref{fig:sloshmdotbh2})} The details of the accretion cycles, {focusing on hydrodynamic evolution for now,} can be seen readily by considering a single cycle in the minidisk around BH2; we choose the cycle achieving peak minidisk mass at $t=52000M$: 
\begin{enumerate}[leftmargin=*,itemsep=0pt,topsep=0pt,parsep=0pt]
    \item Just preceding the mass growth phase of this cycle which starts at $t\sim 51700M$, a sloshing pulse from BH1 strikes the BH2 disk and accelerates accretion of its remaining circularized disk component. Over this period the Reynolds stress
    {  is larger than the
    Maxwell stress, so the Reynolds torque accounts for most of the change} in angular momentum of the disk-like portion of the minidisk {(i.e., excluding the streaming component)}. The incoming sloshing pulse from BH1 results in a {peak for the rate at which Reynolds stress reduces angular momentum (positive plotted value, radial average of $\del_r\{R\}$)}.
    \item Now, with most circularized mass depleted by $t\sim 51800M$, the BH2 minidisk starts to fill with stream material from the close-by lump and remaining received sloshed matter. The accretion rate peaks quickly as the disk mass is replenished. 
    \item The filling of the Hill sphere is rapid, associated with the largest prograde torque integrated across the minidisk {(Reynolds stress contributes the most to angular momentum increase during the circularization process: positive prograde torque 
    {  translates to} $+\del_r\{R\}<0$ in 
    Figure~\ref{fig:sloshmdotbh2})}. As the Hill sphere is filled, the excess becomes partially unbound and forms a sloshing pulse, carrying a significant portion of the angular momentum and energy of the outer minidisk of BH2 across the L1 {Lagrange point} to BH1.
    \item With BH2 now orbiting away from the lump, {stream feeding ends}, and the minidisk mass starts to diminish {at $t\sim 52050$}, due to both accretion and sloshing mass-transfer to BH1 {(note the start of a sloshing pulse seen in the orange curve)}. As the stream feeding cuts off near the beginning of this phase, the minidisk mass is still near its peak, {but the accretion rate drops to near its minimum.} {Thus,
    {  most of the accretion onto a black hole occurs when its minidisk is stream-dominated, and the smallest accretion rates are found in disk-dominated phases, despite the large minidisk mass at those times.}}
    \item As the minidisk around BH2 becomes nearly drained, the cycle repeats.
\end{enumerate}

Although 2D simulations can, at least qualitatively, capture the circumstances in which sloshing is important \citep{Bowen2017,Westernacher2023}, they cannot reveal vertical structure.  At the moment portrayed in Figure~\ref{fig:sloshprofile}, our 3D simulation shows that the dominant sloshing motion is centered on the orbital plane, but is asymmetric with respect to it. At the same time, there is a smaller amount of mass passing the other way, traveling both above and below the dominant sloshing flow. Although at this particular time the sloshing fluxes in either direction are comparable in magnitude, the vertical displacements between the two directions and the consequent strong vertical shear  are generic.   Near the very end of the simulation, when there is nearly continual exchange of material between the black holes, {  a 2D treatment of sloshing mass-transfer could miss important aspects of this flow, like the structure of the shearing layers, revealed only in 3D.} In general, we find that even when the impact of convergent flows occurs predominantly in the equatorial plane around the BHs, out-of-plane asymmetry can form at the convergence point and redirect flows out of the plane. As we will see in the next subsection, sloshing is not the only important new 3D structure discovered.

\subsection{Minidisk tilt}
\label{sec:tilt}
\begin{figure}[]
	\includegraphics[width=\columnwidth]{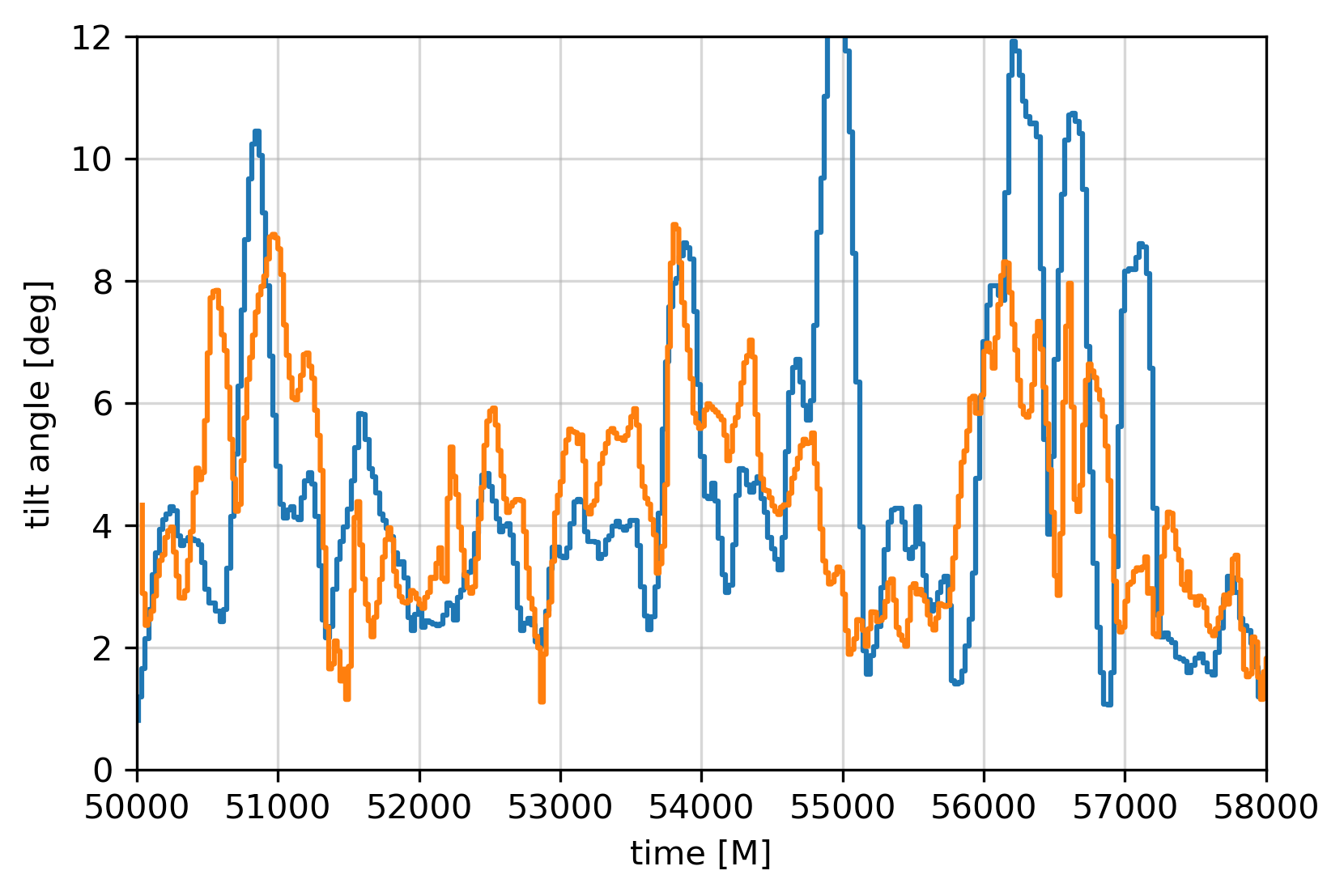}
	\caption{Angular momentum tilt relative to the binary orbital axis: matter entering the Roche lobe of BH1 (blue line) and matter passing inward through $r=21M$ (orange).  The latter has been multiplied by a factor of 2.5.}
	\label{fig:tiltvsmdot}
\end{figure}

In essentially all previous work on minidisks, it has been tacitly assumed that they are aligned with the binary orbital plane.  In these 3D GRMHD simulations, we find that this is almost, but not quite, correct.  If we define the orientation of a minidisk by the orientation of its total angular momentum, the minidisks are, in general, tilted with respect to the binary orbital axis by $\sim 2^\circ - 6^\circ$.  {We have linked this tilt to accreted mass preferentially} leaving the inner edge of the CBD from regions of order a scale height ($\sim 6^\circ$) away from the CBD midplane, {but we have not yet determined {  the origin of} this behavior.} Strong evidence for this connection between tilt and {the dynamics of accretion across the binary cavity} can be found in the close timing relationship between the tilt of the minidisks and the mean tilt of the accretion flow crossing the gap (Fig.~\ref{fig:tiltvsmdot}).  

As a function of time, the azimuthal direction of the minidisk tilt varies with somewhat enhanced occupancy over a range $\sim 100^\circ - 300^\circ$, as measured away from the positive x-axis in the usual way. This twist angle is especially concentrated (by about 50\%) over the narrower range $\sim 150^\circ - 250^\circ$ starting around $t=53000M$ and lasting until around $t=57000M$, when the lump amplitude diminishes.
We speculate that the tilt orientation is related to the apsidal axis of the CBD eccentricity because this is the only non-axisymmetric feature of the system that persists for tens of thousands of $M$ in time, but we leave a quantitative association in angle and time evolution to future studies.

Over long timescales, disk tilt should precess due to angular momentum coupling to the binary angular momentum, but for almost all radii in the minidisks the precession time is much longer than a binary orbit, the residence timescale of matter in a minidisk.

So far, almost every hydrodynamical effect we've explored has little to no temporal coherence on timescales longer than 1.4 binary orbits. This is not true, however, when the global magnetic field evolution is considered.

\section{Magnetic Properties: Flux on the Event Horizon, Magnetic Effects on Fluid Dynamics, and Radial minidisk Angular Momentum Transport}
\label{sec:magnetic}

Magnetic fields influence accretion onto single black holes in several distinct ways: MHD turbulence creates magnetic and hydrodynamic stress that is instrumental in transporting angular momentum \citep{BH98}; Poynting flux carried in buoyant magnetic fields is capable of powering disk coron\ae\ \citep{Galeev1979,NK09,KinchSchnittman2019,KinchSchnittman2021}; poloidal magnetic flux threading the event horizon of a spinning black hole can launch a relativistic jet \citep{BZ77,McKG04,HK06}; and large-scale magnetic fields can support a disk wind \citep{BP82}.  Here we will discuss stress and jet power, and also discover a new magnetic effect specific to binaries.

Our methods give us a number of specific advantages for exploring magnetic effects in the minidisks of accreting binary black holes.  Two in particular are especially important.  First, all the magnetic field brought to the minidisks has its origin in a physically realistic CBD.  Second, our method eliminates the numerous artifacts created by a cut-out surrounding the polar axes and the origin of the coordinate system, in our case coincident with the center of mass.

\subsection{Internal stresses}
\label{subsec:stresses}

The distribution and transport of angular momentum plays a key role in accretion dynamics.  In this relativistic context, the total angular momentum $J$ measured in the prograde orbital direction is the volume integral of the time component of the associated current, $j^\mu$: $J=\int j^t\sqrt{-g}dV$ where $dV$ is the volume element of a spacelike hypersurface.  Here $j^\mu \equiv T^{\mu}_\nu \partial x^\nu/\partial\phi$ for stress tensor $T{^\mu}_\nu$ (Noble12).

We then follow the procedure in \cite{Farris11}, extended to the relativistic context by \citet{Noble12} (Appendix C1), and consider the azimuthal component since that dominates for our relatively aligned system. A few lines of algebra and use of the equation of motion result in the following expression for the time and radial dependence of the spherical shell-integrated angular momentum:
\begin{flalign}
    \del_r\del_t J &= \frac{dT}{dr}-\{\mathcal{F}_\phi\}-\del_r\{T^r_\phi\}\\
                 &= \frac{dT}{dr}-\{\mathcal{F}_\phi\}-\del_r\{M^r_\phi\}-\del_r\{R^r_\phi\}-\del_r\{A^r_\phi\},
\end{flalign}
where $\{X\}\equiv \int \sqrt{-g}X d\theta d\phi$ is the shell integrated quantity $X$, $\mathcal{F}_\phi$ is the rate at which angular momentum is carried away by radiation, $M^r_\phi$ and $R^r_\phi$ are respectively the Maxwell and Reynolds stresses oriented in the $r-\phi$ direction, and $A^r_\phi$ is the advected flux of $\phi$-angular momentum. {$dT/dr$ is the torque density due to purely gravitational effects in the time-varying metric.} The Maxwell stress can be broken down into $M^\mu_\nu=2p_mu^\mu u_\nu + p_m\delta^\mu_\nu-b^\mu b_\nu$, where $p_m$ is the magnetic pressure. This is the electromagnetic part of the stress tensor $T^\mu_\nu$. The hydrodynamic part is $R^\mu_\nu+A^\mu_\nu={T_H}^\mu_\nu=\rho h u^\mu u_\nu + p\delta^\mu_\nu$, with enthalpy $h$ and gas pressure $p$.   Note, however, that $J$ in this context is not rigorously conserved because the spacetime around the individual black holes becomes increasingly non-axisymmetric due to tidal forces as one moves away  from an individual black hole.

As in \citet{Noble12},
we define the non-advected angular momentum flux (the part corresponding to stress in ordinary accretion disks) by the difference between the total angular momentum flux and the advected part:
\begin{flalign}
    \{ R^r_\phi \} &= \{ {T_H}^r_\phi \} - \{ A^r_\phi \}  \\
                  &\simeq \{ {T_H}^r_\phi \} - \frac{   \{-u_\phi \rho /u_t \}\{\rho h u^r \}   }{   \{\rho \}   }.
                  \label{reynolds}
\end{flalign}

There is a subtlety to overcome in the binary context: the angular momentum of interest in the minidisks is defined relative to their individual polar axes.   To find the flux of this quantity, rather than the angular momentum relative to the coordinate polar axis,
we first transform the quantities on the central patch from internal numerical coordinates to physical Cartesian coordinates. We then coordinate transform the 4-velocity $u^\mu$ and magnetic 4-vector $b^\mu$ to bring them into the instantaneously translating frame of the BH of interest.   The final transformation changes the Cartesian coordinates in this frame to spherical coordinates centered on the BH, and then for numerical ease in analysis of, for instance, shell integrated quantities, we interpolate into spherical grids around each BH. The resolution of the final grid has been found by convergence testing to reach a threshold of $\lesssim5\%$ error at most in quantities of interest.

The shell-integrated torques due to the Maxwell and Reynolds stresses are the negatives of the radial derivatives of $\{M\}$ and $\{R\}$ respectively. Much of the time these derivatives have little net trend (although with large variance) between $r=4$ and $r=8$, justifying a radial average across this range to get an instantaneous value for the torque on minidisk gas. The torques on minidisk material for each BH Hill sphere as functions of time are shown in the {middle} panel of Figure~\ref{fig:sloshmdotbh2}  (note that negative values of this gradient correspond to a net gain of angular momentum, i.e.,  a positive torque).

The hydrodynamic stress (and resulting torque) nearly always dominates over the magnetic stress and its torque. 
When minidisks accrete, angular momentum is transported outward, but the net hydrodynamic torque $-\partial_r\{R\}$ is often positive as the angular momentum at a given radius grows with the orbiting mass at that location. The peak positive hydrodynamic torque coincides with the transfer of angular momentum into sloshing material and expulsion of this gas to the other minidisk. 

The previous paragraph spoke of ``net torque" because, unlike traditional MRI-dominated disks, there are numerous large-scale coherent structures in the minidisk in which negative and positive torques are juxtaposed.   They  are variously compression fronts, standing spiral shocks, and sloshing-induced shocks.   Because there is a close balance between positive and negative torques, the time-averaged ratio  $\langle|\{R\}+\{M\}|/|\{A\}|\rangle\sim0.1$, indicating that advection dominates the net transport of angular momentum. This is not surprising because such a large fraction of the material either streams directly into the BH or sloshes between the BHs. 

Although the hydrodynamic torque contribution varies significantly based on the minidisk accretion state, torques from magnetic stress, on the other hand, {closely} track the total mass of the minidisk.  At later times, the hydrodynamic torque may behave more like the magnetic torque.
Once the binary becomes very close, the state cycles are no longer clearly defined; instead, minidisks remain primarily in the stream-dominated state.  When this is the case, minidisk structure varies little over time, so the distinction between the two stresses is erased.

\subsection{Magnetic flux on the horizon}
\label{subsec:horizonflux}

A useful measure of dimensionless magnetic flux on the horizon that gives insight into potential jet launching is the commonly considered (especially in literature related to Magnetically Arrested Disks \citep{IgumenshchevNarayan2003,NarayanIgumenshchev2003}), $\phi$. The definition of $\phi$ begins by first taking the integral of ``absolute flux", 
\begin{equation}
\Phi(r) \equiv \frac{1}{2}\int  r^2 d\Omega |B^r|,
\end{equation}
where $\Omega$ is solid angle and the integral is taken over the complete spherical surface. Because the Blandford-Zjajek (BZ77) process is locally symmetric with respect to sign of the magnetic field, $|B^r|$ is more relevant than $B^r$.  The $1/2$ makes it more directly comparable to the integral over the signed field, which is conventionally integrated over a hemisphere.
$\Phi$ is usually measured on the horizon, but the effective resolution for interpolation into spherical coordinates $\Delta x/r$ improves with distance $r$ from the black hole.  We therefore integrate magnetic fluxes just inside the ISCO. \footnote{This offset is also often considered in related single-BH studies to avoid calculation of $\dot{M}$ at the horizon where floor matter injection can be significant, but this is not an issue in our case since we evolve without spin on the BHs.}

The quantity most useful for relating horizon-scale magnetic flux to potential jet power if the BHs had significant spin (\cite{Tchekhovskoy2014}, \cite{Avara2016}, etc) is $\Phi$ normalized by the square root of the accretion rate
\begin{equation}
    \phi = \frac{\sqrt{4\pi}\Phi}{\sqrt{\dot M}}.
\end{equation}
This can be related to the dimensionless flux measure of \cite{G99}, $\Upsilon\approx0.2\phi$. In some papers the normalization of these quantities is by the time-averaged accretion rate $[\dot M]_t$, but in all cases here we use the \cite{Tchekhovskoy2014} approach and normalize by the instantaneous value which, in inflow equilibrium, has been shown to be just as robust.

On the other hand, to analyze long-term behavior of horizon-scale magnetization, we also consider the complementary signed integral of flux,
\begin{equation}
\Psi_{\rm (up,dn)}(r)  \equiv 
\int_S \, r^2 d\Omega B^r, \\
\end{equation}
where the surface $S$ is the upper (lower) hemisphere for $\Psi_{\rm up,(dn)}$.
This quantity can be used to provide insight into the global evolution of magnetic flux. The ratio of $|\Psi|$ to the same integral over $|B^r|$, $|\Psi|/\Phi$ quantifies the importance of substructure and can be related directly to azimuthal components of the spherical harmonic decomposition of $B^r$ over the surface at $r$ \citep{McKinney2012}.

\begin{figure}[]
	\includegraphics[width=\columnwidth]{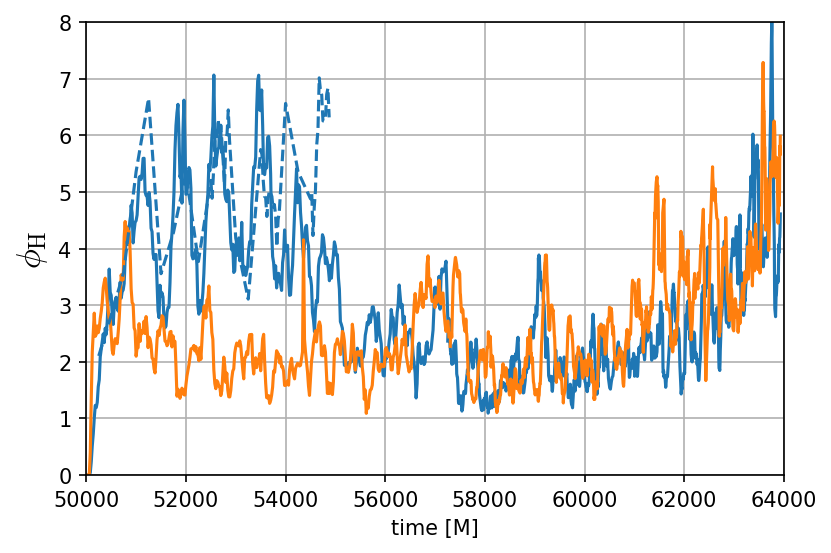}
	\caption{Time-dependence of $\phi_{\text{H}}$, the normalized absolute magnetic flux on the ISCO of each of the black holes (BH1 blue, BH2 orange lines), for PM.IN20s.  Values from PM.IN20sHR for BH1 are shown with blue-dashed line; they are generally close to the fiducial resolution values. }
	\label{fig:absflux}
\end{figure}

\begin{figure}[]
	\includegraphics[width=\columnwidth]{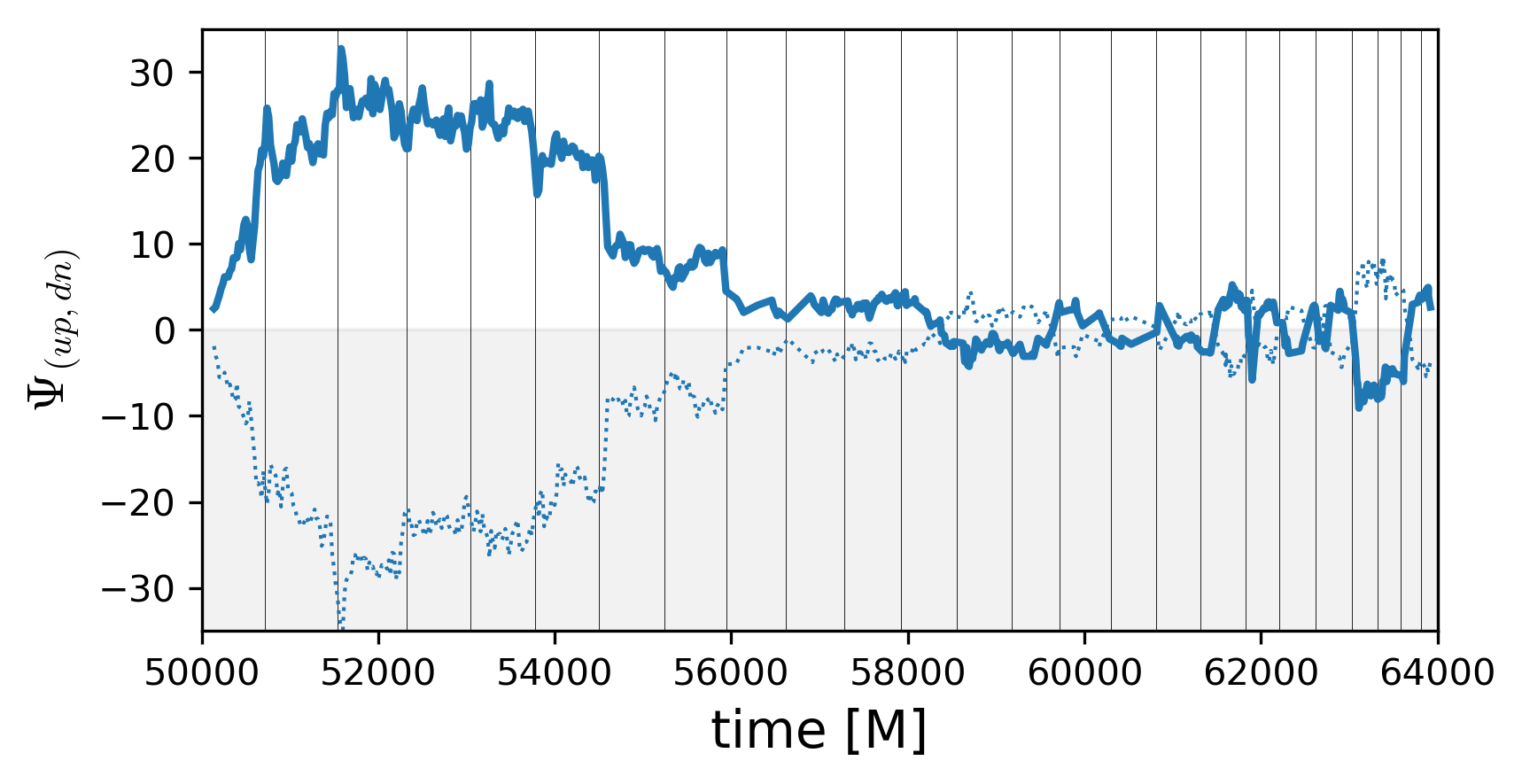}\\
	\includegraphics[width=\columnwidth]{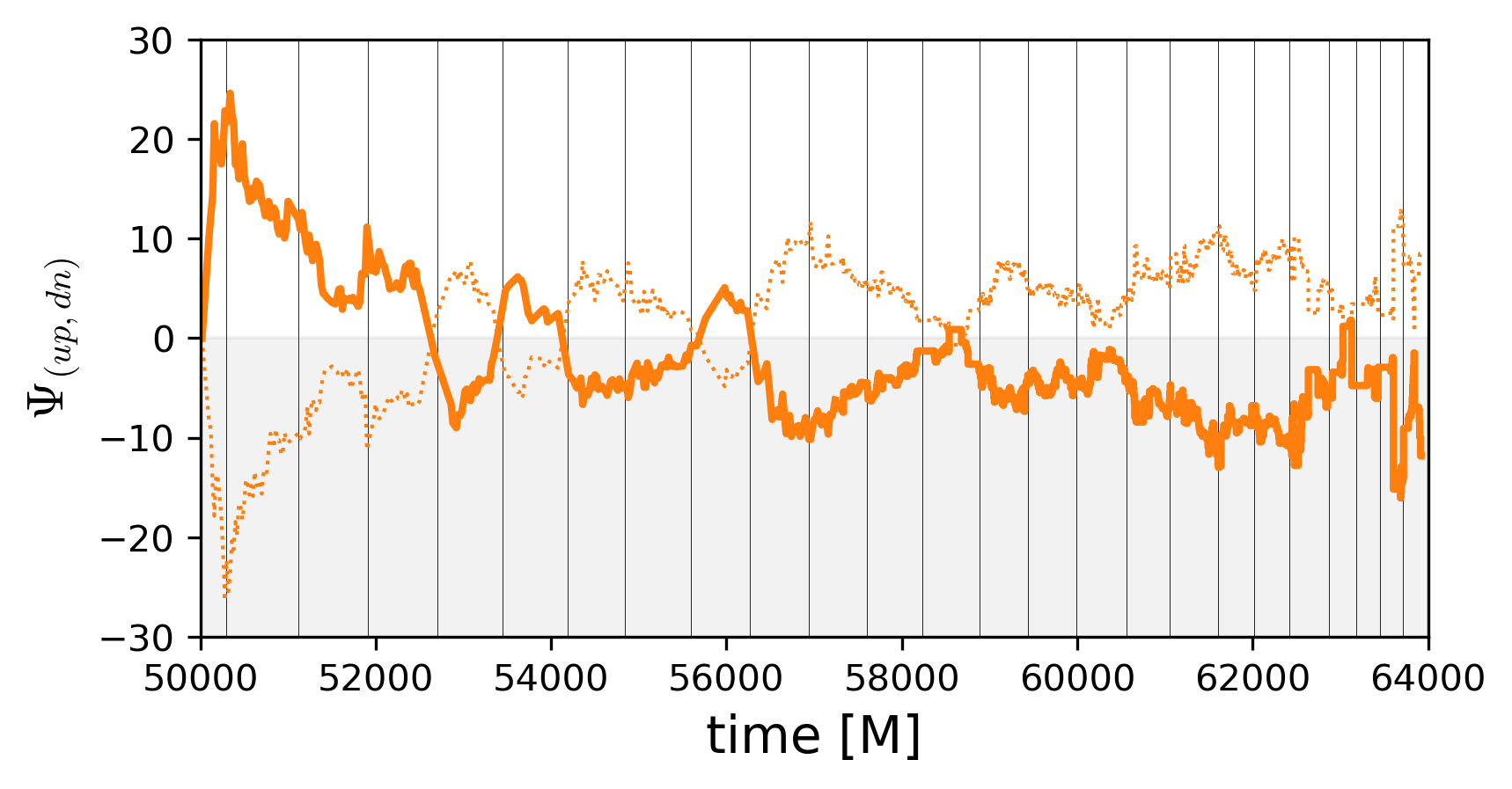}
	\caption{Upper (solid-thick) and lower (dotted-thin) hemisphere-integrated signed magnetic flux $\Psi$. The upper panel is for BH1, the lower panel for BH2. Vertical-black lines indicate times each orbit where the BHs share orbital phase with the centroid of the lump, i.e. the times of closest approach.}
	\label{fig:signedflux}
\end{figure}

The time-dependence of $\phi$ is presented in Fig.~\ref{fig:absflux}. 
Unfortunately, the first $\lesssim 5000M$ for BH1 appear to be marred by an artifact of our initialization procedure arising from a coincidence in which a peak in magnetic field in the data we took from Noble12 led to an anomalously large magnetic field on BH1 at early times in the simulation; the value of $\phi$ on BH1 during this transient should therefore be ignored.
After $\approx 55000M$, the magnitudes of $\phi$ on the two black holes evolve almost in lock-step.  Between that time and $\approx 61000M$, $\phi$ for both remains $\simeq 2$ except for occasional brief peaks at $\simeq 4$.  However, after $\approx 61000M$, $\phi$ for both rises steadily, reaching a value $\simeq 5 -6$ at the end of the simulation.  Focusing on BH2 in order to avoid the transient artifacts in $\phi(t)$ for BH1, it is clear that $\phi$ varies less than the accretion rate (cf. Fig.~\ref{fig:MdotandRunSE}) on near-orbital timescales.  It is not, however, devoid of fluctuations; in particular, there is a weak correlation between peaks in $\phi$ and pulses in the accretion rate.

The signed fluxes $\Psi_{\rm up,dn}$ also exhibit a short-timescale correlation with \dotm, identified in Fig. \ref{fig:signedflux} with vertical lines indicating each close passage of the BH by the lump. 
Though most sharp changes in $\Psi$ can be associated with onset of an accretion episode, the reverse is not true for all accretion pulses. This would be expected if accretion pulses bring in material with a range of magnetizations.
In addition, if magnetic flux is continually brought to the black hole horizon and kept there, at least in part, by the ram pressure of the accretion flow, both overall smoother variation and correlation with recently-completed accretion pulses might be expected.

If the black holes had significant spin, they might both support jets.  Studies of accreting single black holes find that the time-averaged jet power is an increasing function of the black hole spin parameter times $\phi^2$ (see \cite{DavisTchekhovskoy2020} and references therein).
Moderately thin simulations of MAD disks \citep{Avara2016,TeixeiraAvara2018,Liska2022a} show that at values of $\phi$ ranging from 25-45, black holes with spin parameters $\approx 0.5$ \citep{Avara2016,TeixeiraAvara2018} and and near maximal \citep{Liska2022a} can power jets with sizable rest-mass efficiency ($\approx 50\%$ in the latter case).  Because this value of $\phi$ is about a factor of 10 greater than the black holes possess for most of the duration of our simulation, an efficiency scaling  $\propto \phi^2$ predicts a jet efficiency reduction for the black holes by a factor of $\sim 10^{-2}$ from the MAD efficiency, to $\sim 0.005$.   This may rise by a factor of several as the separation falls to $\lesssim 10M$ and $\phi$ rises to $\simeq 5$.

\subsection{Relation to accretion states}
\label{sec:magstates}

\begin{figure}[]
  \begin{centering}
   \subfloat[Disk dominated: $t=56025M$]{
	\includegraphics[width=\columnwidth]{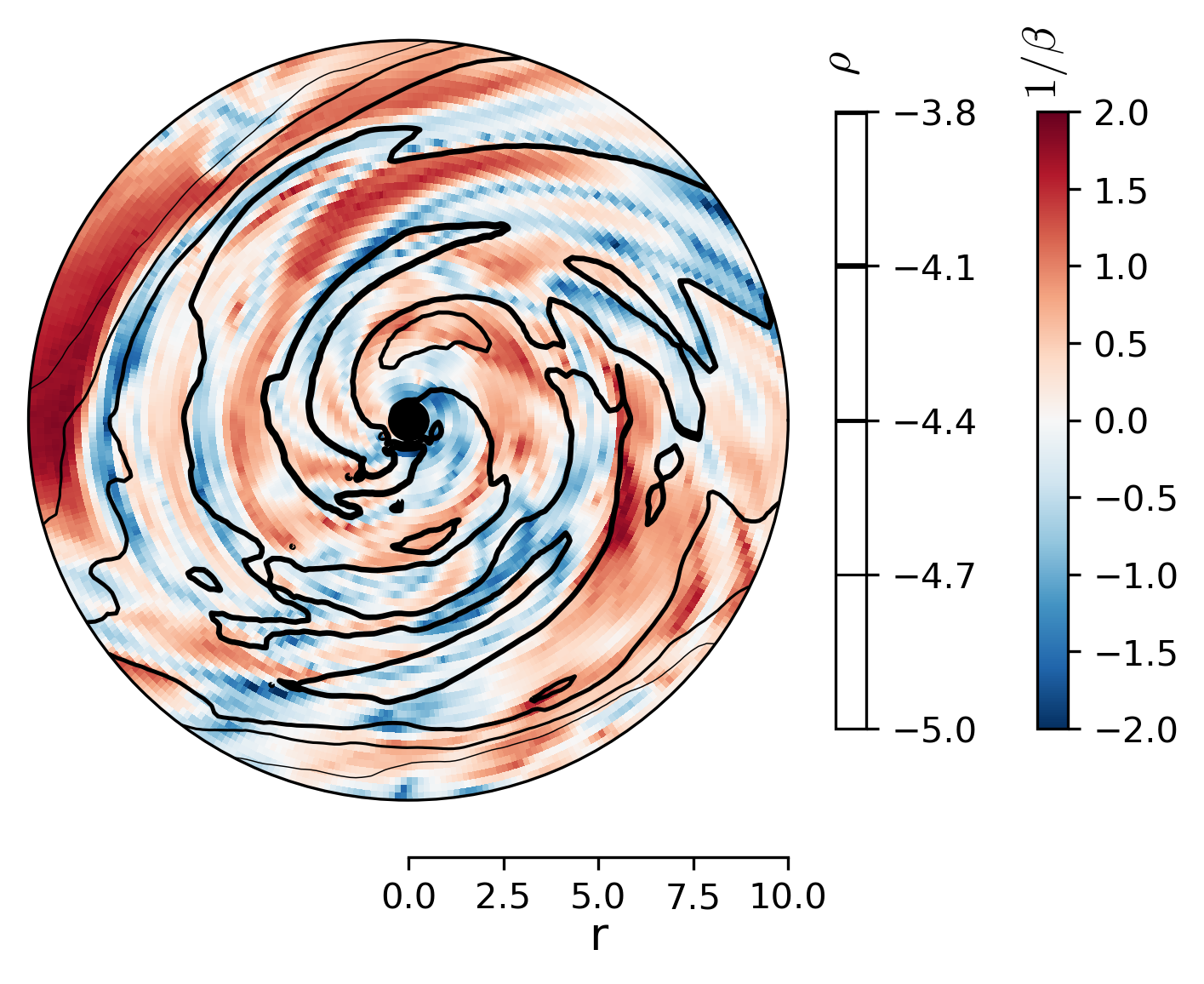}   
	}\\
	  \subfloat[Stream dominated: $t=55600M$]{	
	\includegraphics[width=\columnwidth]{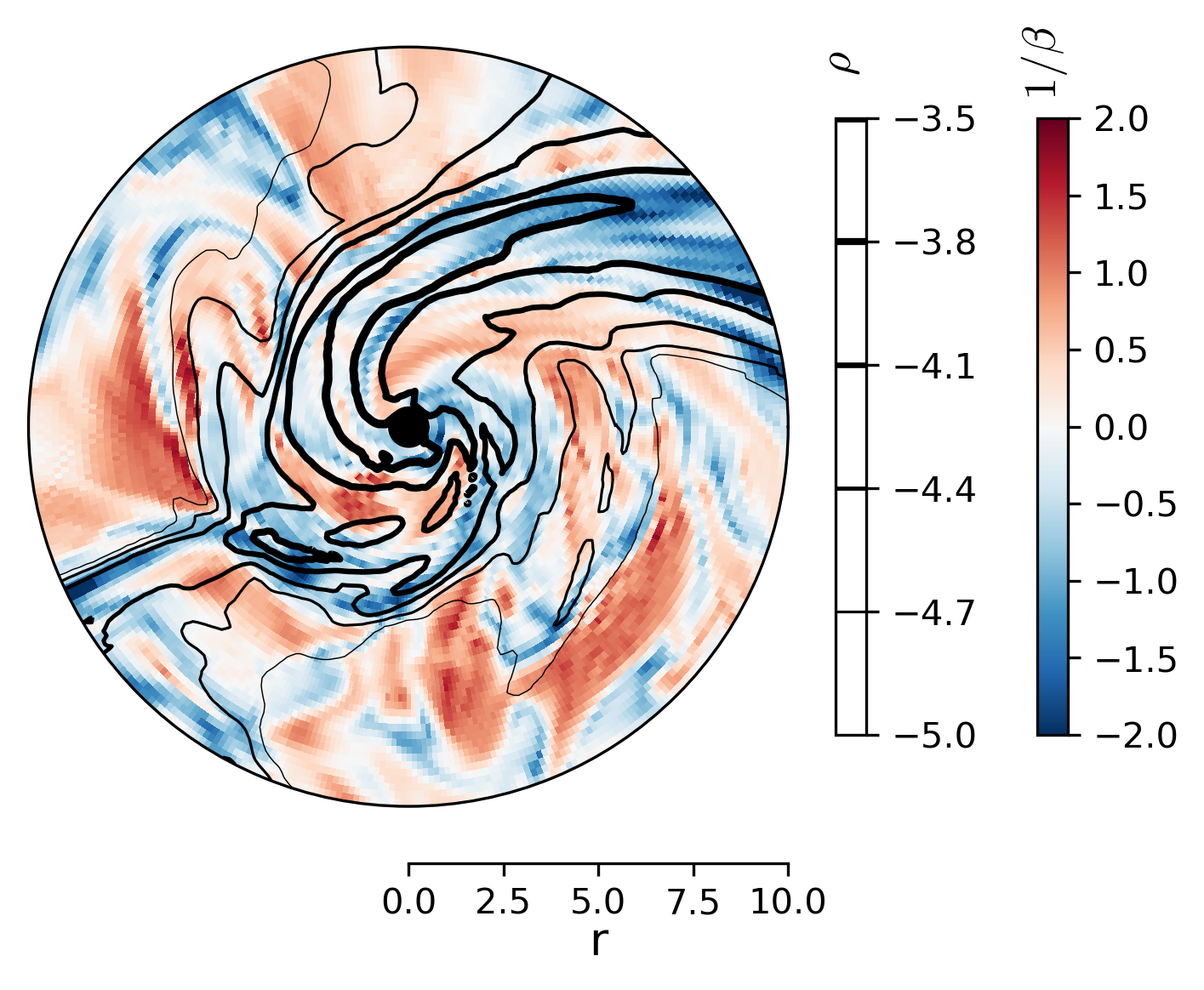}
	 }
          \caption{Like Fig.~\ref{fig:strmvsdisk}, but with $\mbox{log}_{10}(1/\beta)$ in color-scale. Note the striping of $\beta^{-1}$ values inside the stream where the magnetic flux is much larger in magnitude than the disk-like portion of the flow, and stretched along the stream. This is where the Maxwell stress dominates in the streaming minidisk state. 
	}
	\label{fig:inversebeta}
  \end{centering}
\end{figure}

As previously discussed in \S\ref{sec:accretionstates} and summarized in \S\ref{subsec:sumcyc}, the short residence time of matter in the minidisks leads to the disks alternating between two different states, disk-dominated and stream-dominated. The character of the magnetic field also changes sharply between these two states.  As shown in Figure~\ref{fig:inversebeta}, the disk-dominated state is mostly gas pressure-dominated (plasma $\beta \equiv p_{\rm gas}/p_{\rm mag} > 1$), whereas the stream-dominated state is mostly magnetically dominated ($\beta \lesssim 1$).

Because the {\it ad hoc} cooling function we invoke targets a specific temperature regardless of the minidisk state, in the magnetically dominated phase, the low-density portion of the streaming state minidisk should have a scale height to radius ratio $H/r_{BH}$ close to the target value of 0.1 if the gas has sufficient time to cool. In the frame of a BH, the cooling time across the minidisk varies across the range {$\Delta t_{cool}\sim 80 - 250M$} between the ISCO and the tidal truncation edge. We find that the inflow time of the disk portion of the flow, $r/<v_r>$  time averaged over all times/disk-states, spans a similar range for the same radii.  As a result, material in the disk-like portion of the flow has just barely enough time to cool. In practice, the entropy is almost twice the target value, and $H/r_{BH}$ rises from $\sim 0.1$ to $\sim 0.2$ running outward.  The stream component is much thicker, with $H/r_{BH}$ between 0.4 and 0.5. In the stream, the inflow time is also significantly shorter than a cooling time.
During stream-dominated states, when the disk part of the flow is magnetically dominated, magnetic pressure supports the disk vertically.


Unlike ordinary  accretion disks, the magnetorotational instability (MRI, \cite{BH91}) does not play an important role.  As we have already seen, the Maxwell stress is generally smaller than the Reynolds stress, contrary  to the prevailing situation of MRI-stirred MHD turbulence.   This is because the disk residence time is short, only $\lesssim 1/2$ binary orbital period, or $\sim 2$ orbital periods at the outer edge of a minidisk.   This is much shorter than the usual nonlinear saturation timescale of the turbulence, $\sim 10$ local orbital periods.

\label{sec:lightcurve}

\section{Discussion}
\label{sec:discussion}

\subsection{A key methodological advance}

This simulation would have been prohibitively expensive without the use of our \pwmhd\ infrastructure.  By evolving the flow independently on two different grids, which combine to cover the entire problem domain, we were able to cut the computing time by a factor $\simeq 30$.  This saving is achieved by the freedom this method gives to choose grids optimized for their regions of use; by this means, we avoid cut-outs around coordinate singularities, achieve resolution quality that would require many more cells on a unitary grid, and need many fewer time-steps to evolve over a given physical time.

\subsection{The nature of ``decoupling"}

Since the work of \cite{MP05}, it has been recognized that there comes a point in the evolution of an accreting binary black hole system at which, due to energy-loss by gravitational wave radiation, the orbit shrinks on a shorter timescale than the timescale on which internal stresses inside the CBD cause gas to migrate inward.  Initially this timescale contrast was interpreted as signalling an end to accretion from the CBD to the minidisks.  However, since the work of \cite{Noble12}, \cite{Farris2015}, and \cite{Tang2018}, it has increasingly been seen as a point at which the accretion rate onto the minidisks diminishes, but by a factor of order unity that may be sensitive to the binary parameters and the specific state of the CBD.

By following the inspiral from $20M$ to $\simeq 9M$, and including evolution of the cavity as a more realistic inner boundary condition for the CBD evolution, we have been able to describe this process in greater quantitative detail than before.
During the first half of the simulation, in which the binary shrinks from $20M$ to $\approx 17M$, the rate at which mass reaches the minidisks falls by a factor $\sim 4$, but is roughly constant from then until the end of our simulation.
At the same time, however, the accretion rate outside the lump/over-density region of the CBD remains at roughly its initial value.
During this time, the lump region of the CBD departs from the behavior inherited from RunSE. In RunSE at the time we start PM.IN20s the qualitative structure of the inner CBD was well established, but a small imbalance between the material accreting into the lump region and the material flowing past it and into the binary cavity caused a slow growth of surface density in the lump region. This is evident in the region between $r=35$ and $55M$ in Fig. \ref{fig:mdotvsrfull}.  As inspiral accelerates, the lump growth continues, but the decoupling process begins as well.

These results point toward two gaps in the initial analysis of decoupling.  The first is that if the ratio between black hole accretion rate and CBD accretion rate decreased with a decreasing ratio of binary orbital evolution time to CBD inflow time, one would expect the accretion rate ratio to fall steadily---but instead it initially falls and then reaches a near steady-state.  The second is more fundamental.  To support the full accretion rate for the time remaining before the black holes merge requires an amount of mass at the inner edge of the CBD that is only $\dot M_0 t_{\rm evolve}$.  By the definition of ``decoupling", this is a small amount of mass compared to the disk mass in its innermost $e$-fold in radius.  When mass travels from the inner edge of the CBD to the minidisks, it follows a ballistic orbit, which can be traversed in time comparable to a binary orbit, which is even shorter than $t_{\rm evolve}$.  What dynamical process prevents this small amount of mass from quickly joining the binary?

We suggest two answers to these questions.  First, the continued accretion onto the minidisks, albeit at a lower rate, may be due to an intrinsic radial gradient in the internal stress at the inner edge of the CBD.  Although the ratio of Maxwell stress to pressure within the disk body varies little as a function of radius, it increases sharply at the inner edge \citep{Noble12,Shi12}.  Thus, although the orbital evolution may be faster than the bulk inflow rate, it may not be faster than the inflow rate within the surface density peak at the CBD's inner edge.   Second, the time-dependent portion of the binary's quadrupole moment breaks the axisymmetry of the gravitational potential, creating orbits on which the polar component of angular momentum is not conserved.  It is by this means that matter can quickly fall from the inner edge of the CBD to the minidisks.  This quadrupole moment decreases $\propto a^2$ as the orbit shrinks.
One consequence is that the processes reinforcing the lump structure diminish, and the lump stretches in azimuth. 
Another is that the inner edge of the CBD becomes less distinct, more uniformly distributed in azimuth, and contains more mass between the original CBD tidal truncation edge and the shrinking binary, a new region allowing quasi-circular orbital stability. 

\subsection{Minidisks as unconventional accretion disks}

Accretion disks are generally thought of as nearly axisymmetric objects fluctuating around a slowly-varying state of nearly equilibrium inflow.  The dominant mechanism promoting inflow is thought to be Maxwell stress due to correlated MHD turbulence driven by the magnetorotational instability.  This view of the minidisks is generally viewed to be appropriate for binaries with larger separations.  However, when the binary separation is not a great deal larger than the size of the ISCO orbits around the black holes, the work of \cite{Gold2014a}, \cite{Bowen2018,Bowen2019}, and \cite{Combi2021} has shown that the minidisks operate very differently from the classic picture, and we have extended our understanding of just how unconventional minidisks in this regime are.

Most strikingly, the minidisks are anything but slowly-varying and axisymmetric.  They flip back-and-forth between a low-density stream-dominated state in which nearly all their mass is concentrated in a narrow stream plunging toward the black hole, and a disk-dominated state in which their mass is distributed much more broadly in azimuth, but is nonetheless far from axisymmetric, having little time to blend azimuthal structure before accreting.  In this latter state, although gas orbits a number of times before reaching the ISCO, its mean inward speed is nonetheless a considerably larger fraction of the orbital speed than in a conventional accretion disk.  Moreover, the forces acting on the gas in these disks are predominantly gravity and Reynolds stress associated with spiral features.  Maxwell stress is a secondary effect even in the streams, despite compression diminishing the plasma $\beta$ somewhat.

The heating that powers radiation is therefore due much more to fluid dissipation mechanisms than magnetic dissipation.  Because the two disks' phases with respect to the accretion state oscillation are opposite, the total heating rate is relatively steady despite the nonlinear modulation of each one individually.  Moreover, the close binary separation implies moderately relativistic orbital speeds for the black holes, so the luminosity from the disks should be strongly modulated on the orbital frequency by Doppler effects.  However, this is not the only periodic modulation: the disk-feeding oscillation period is {longer} than the orbital period.  The power spectrum of the lightcurve should therefore exhibit peaks at both periods and exhibit significant sub-structure related to the detail accretion and emission behavior.

\subsection{Sloshing {in 3D}}

As we have shown, the typical rate of mass transfer between the minidisks is substantial: $\sim 0.1\times$ the total accretion rate.  Due to the symmetry in our equal-mass simulation, there cannot be any long-term net transfer.  However, in a binary with $q \neq 1$, we expect there will be a systematic trend in the direction of the minidisk mass-exchange.  If its magnitude is comparable to the total sloshing mass-transfer rate, it could alter how rapidly accretion tends to bring the black holes closer to equal mass, with attendant consequences to both accretion dynamics and the rate of gravitational radiation. 

The complex 3D structure of the sloshing and its association with minidisk tilt underscores the necessity of 3D simulations to correctly capture this effect on mass ratio evolution.

\begin{figure*}[ht]
	\includegraphics[trim={0cm 0cm 0cm 0cm},clip,width=1.0\linewidth]{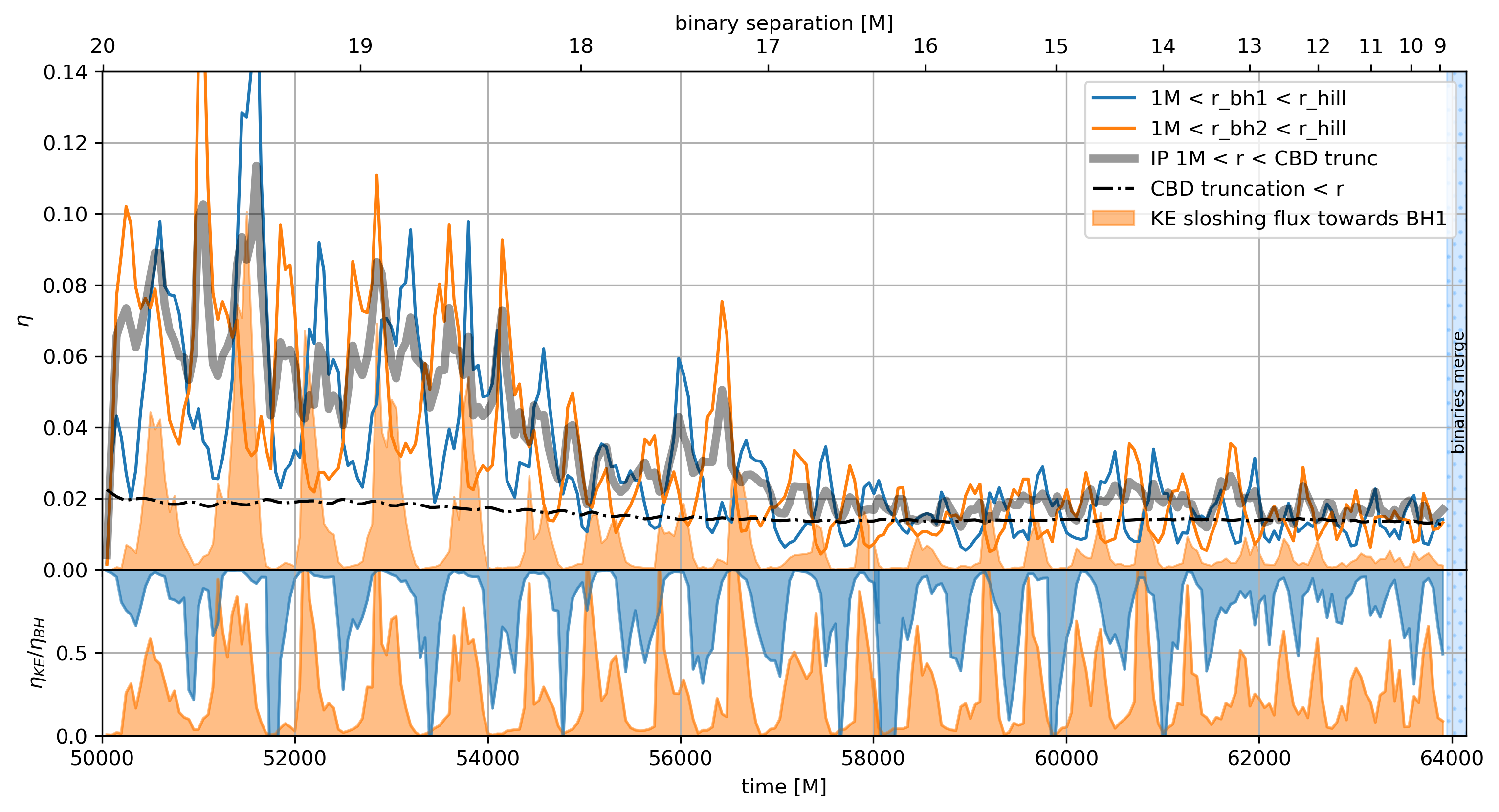}\\
	\caption{{\bf [Top Panel]} Integrated energy dissipated in the cooling function in different regions of the domain, plotted as an efficiency fraction $\eta$ as normalized by the mass accretion rate. See the text for a discussion of the time-averaged accretion rates chosen. Blue and orange lines are integrated over the Hill spheres of BH1 and BH2 respectively. The integral is made over the full sphere between 1M and 8M$\times$(20/a(t)), scaled with the separation as the binary inspirals. Gray-thick line is efficiency for the integrated cooling function over the entire cavity outside of a 1M radius around each BH and inside the CBD truncation edge, so including both minidisks and the low-density cavity/streaming region, normalized by the combined accretion rate onto both black holes.  Black-dash-dot line is the integral over the CBD. Orange-shaded region falls under the line indicating the kinetic energy in the sloshing flow originating from BH2, normalized by \dotm of the receiving BH, so as to have comparable units to the radiative efficiency. {\bf [Bottom Panel]} In blue and orange we plot the KE efficiency of sloshing as a fraction of the radiative efficiency of the receiving BH. For ease of comparison, for sloshing flux originating from BH1 (indicated in blue) we plot $(1-\eta_{\rm KE}/\eta_{\rm BH1})$. }
\label{fig:coolfuncvstselect}
\end{figure*}

In addition to the transfer of mass flux, the sloshing carries significant kinetic energy, which can be dissipated in the shocks formed when the sloshing flow strikes the recipient minidisk.  At times of peak sloshing rate, the dissipation of this energy contributes substantially to the total bolometric luminosity of the system.  If the kinetic energy is dissipated and radiated promptly, it can briefly dominate the radiative output of the receiving minidisk (see the bottom panel of Fig. \ref{fig:coolfuncvstselect}).  In a time-averaged sense, it could also increase the radiative efficiency of the flow by leaving the gas with low enough orbital energy that it crosses the ISCO with less than the energy that would support a circular orbit there. 

\subsection{Magnetic flux on the black hole event horizons}

In the underlying {$t=0$ initial conditions} of our simulation, i.e. {the initial conditions of RunSE of \cite{Noble12}} {from which we pick up evolution at $t=50000M$ after adding the full binary cavity}, the magnetic field was given a very simple structure: nested dipolar field loops following the density contours of the initial CBD. {The field  amplitude was chosen }such that the ratio of the total fluid internal energy to the total magnetic energy was 100 everywhere the density was above a low threshold value, that of the torus's edge. From that point on, the magnetic field's geometry and intensity were determined entirely by the operation of ideal MHD in the evolving binary spacetime.  These physical processes amplified the field in the CBD everywhere but the lump that eventually forms, and this field was then conveyed to the minidisks and their black holes by physically-determined motions.  Nonetheless, in the relatively brief duration of our simulation, $\simeq 14000M$, the magnetic flux on the black hole horizons grew to a level such that, if the black holes had significant spin, a pair of relativistic jets would be launched whose ratio of power to accreted rest-mass is comparable to the minidisks' radiative efficiency by the end of the simulation, and a significant fraction at all earlier times. We can therefore expect that in typical cases jet power will be comparable to photon radiation. However, the specific ratio depends strongly on the spin parameters of the individual black holes, and the absolute jet power depends on the accretion rate and magnetic flux delivered to the spinning black holes.

Interestingly, we find that the net integrated flux threading the BHs varies on a super-orbital timescale, with no clear cyclical behavior in either magnitude or sign. The lack of evidence for modulation of horizon scale flux associated with the minidisk state cycle suggests that the large scale magnetic flux available to the black holes is instead driven by CBD behavior, at least through these final stages of inspiral. 
By comparing the behavior of $|\Psi|/\Phi$ to its constituent parts, though this estimate is much more crude than full analysis of the Fourier decomposition, we find that the higher multipole contributions fluctuate following the minidisk accretion cycles.
Because higher multipole components in poloidal field appear to contribute more weakly to jet power \citep{BHK08} these fluctuations could be an additional source of variability to any jets.

Lastly, we speculate that the horizon-scale magnetization we report, and the associated potential for jet power, may be a conservative estimate. Although the trend we see does not suggest a simple secular evolution, evolving the binary from wider initial separation could possibly give the BHs time to accumulate more magnetic flux before entering this final inspiral regime, possibly resulting in higher final magnetic flux.

\subsection{Minidisk tilts}

For the first time, with a careful analysis of the angular momentum and matter distribution in the minidisks, and the angular and time dependent profile of mass accretion into the cavity, we have found tilting minidisk behavior that is driven by accretion streams preferentially originating from regions vertically offset from the midplane. Though the long-time average accretion rate is vertically symmetric across the midplane, each minidisk cycle is dominated by material with a vertical {break in symmetry}. Analysis of vertical slices through the regions where streams interact with the CBD truncation edge reveal that this behavior could be in part due to vertical expulsion of material due to convergence of flow in the midplane. It is also possible that the CBDs exhibit an intrinsic vertical accretion profile that differs from the standard picture. Preferential accretion along the surface of disks has been seen in simulations of single-BH accretion but is generally associated with large-scale poloidal magnetic flux threading the disk \citep{BHK09,Avara2016}.

No matter the detailed origin of the asymmetry in the episodic accretion, tilts would likely affect the photon flux received by distant observers in two ways: 1) they change the projected area of the minidisks as seen by these observers, as well as moving such observers to different angles in the limb-darkening function of the minidisks' surfaces; 2) they alter the angle between the orbital velocity and an observer's line-of-sight, thereby affecting Doppler boosting and beaming.  These effects may also depend on the scale heights for the minidisks and CBD, as well as on the vertical profiles of the shocks where accretion streams from the CBDs strike the minidisks.

\section{Conclusions}
\label{sec:conc}

We have reported run details and analysis of 3D-GRMHD global simulations of an inspiraling SMBBH as it evolves from 20M to 9M separation. This constitutes 98\% of the temporal evolution before merger from this initial separation, with equal mass BHs in a quasi-circular orbit. This simulation is accelerated and achieves greater physical realism than prior comparable runs by using the \pwmhd+\harmd code infrastructure; this is the first known application of the multi-mesh methodology to simulations of binary accretion. 

Our primary findings fit roughly into two categories, those pertaining to hydrodynamic evolution and energy dissipation and those related to global magnetic field evolution.  Four hydrodynamic effects are of particular interest.  First, the minidisks during the last phase of inspiral are very  different from ordinary accretion disks. They cycle between a stream-like and a disk-like state, and  in neither state does Maxwell stress dominate Reynolds stress.  Second, the binary sustains significant rates of mass accretion and energy dissipation throughout the inspiral period.  Third, inter-minidisk ``sloshing" (previously noted by \citet{Bowen2017}) carries as much as $\sim 10-20\%$ of the instantaneous accretion rate from one minidisk to the other; it can also transport kinetic energy at a rate comparable to the minidisks' luminosity.
Fourth, the flow exhibits significant 3D geometry.  The minidisks are tilted relative to the orbital plane, and this tilt changes over time.  In addition, the sloshing flow is divided into multiple shearing layers, with most mass flux often traveling in one direction in the midplane and in the opposite direction above and below the midplane.

{Each of these findings may have significant observational consequences. As the system traverses the cycle of minidisk states, it is likely that significant variability in 3D structure, density, magnetization, and shock-heating will lead to significant differences in dissipation properties and spectral signature. For instance, the low density and higher magnetization of the disk component of the stream-like state might lead to a harder spectrum emitted from most of the minidisk surface area, potentially contrasting with a thermal component from the dense stream. It is possible that detailed treatment of the radiation in this late phase of the binary inspiral might reveal significant and complex periodic structure.} 

{Moreover, sloshing does not exist in single-BH systems and becomes especially strong only in the final inspiral.  In that stage, we have shown it can contribute significantly to the bolometric luminosity, although the band in which sloshing's contribution is most visible has yet to be determined.}  {Finally, any geometric change to the minidisks or CBD can potentially have a measurable impact on the overall EM properties:} {  through flux modulation due to varying projected area, through line profile modulation, and} {by affecting column depth along line-of-sight to the observer.}

{Beyond cyclical behavior, by extending our 
simulations longer than prior comparable models, we discover that as the binary approaches merger, the lump dissipates and stretches in azimuth, some CBD accretion decoupling occurs, and these balancing processes result in a relatively modest reduction of the accretion rate compared to analytic predictions.
}We find that the total CBD contribution to total accretion efficiency is $\sim2\%${.  {  The bolometric lightcurve of the cavity containing both minidisks and streams but excluding any jet  contribution, is the convolution of the accretion rate with the radiative efficiency, which falls gradually from comparable to that of single black hole thin-disk theory to only slightly more than that of the CBD.  Consequently, the cavity and CBD contributions are comparable in bolometric luminosity, but distinguishable by  sharp contrasts in emission and band and time-dependence.} 
All these effects could have important implications for the observability of merging SMBBHs. }

Just as in accretion onto single black holes, accretion {regulates }magnetic field strength and structure on the horizon(s). By starting from a pre-equilibrated CBD evolved in 3D-GRMHD, and then evolving the global binary system, we have, for the first time, directly connected the large-scale turbulent magnetic field in a quasi-equilibrium state of a CBD to the magnetic field on the much smaller horizon scale. {Unlike minidisk behavior, which varies on the orbital timescale, we find the dominant time-variability of the magnetic field both in the minidisks and on the black holes to be slower, 
{  due to the accumulation of the flux transported from the CBD or possibly to magnetic evolution of the CBD.} 
This result has important implications for how variability of jet-associated EM emission is driven in late-stage binary evolution. Lastly, we confirm the potential for significant jet power.} Even during the comparatively short period of our simulation, the poloidal flux on the horizon grew strong enough that if the black holes were spinning rapidly the jet power would reach $\sim 0.2\%$ in accreted rest-mass efficiency for most of the inspiral, and possibly $\gtrsim 1\%$ efficiency immediately before merger.

\begin{acknowledgments}
We thank Vassilios Mewes, Hotaka Shiokawa, and Roseanne Cheng for insightful discussions related to PatchworkMHD that ultimately improved the work described in this manuscript. M.J.A. additionally thanks Daniel D'Orazio, Zoltan Haiman, Jonathan Zrake, Andrew MacFadyen, Alessia Franchini, Dong Lai, Luciano Combi, Federico Lopez-Armengol, and Chris Reynolds, for valuable scientific discussions.

This work was also supported by several National Science Foundation (NSF) grants.

M.J.A was supported by a "Frontiers in Gravitational-Wave Astronomy" Fellowship of the RIT's Center for Computational Relativity and Gravitation, as well as support from a grant of the European Research Council under the European Union’s Horizon 2020 research and innovation programme (grant 834203).

M.J.A and M.C. gratefully acknowledge support from NSF awards NSF AST-2009330, AST-1516150, PHY-2110338 and PHY-1707946. 
J.H.K. received support from AST-2009260, AST-1516299, PHYS-2110339 and PHY-1707826.

S.C.N. was supported by an appointment to the NASA Postdoctoral Program at the Goddard Space Flight Center administrated by USRA through a contract with NASA.

Computational resources were provided by the NCSA's Blue Waters supercomputer allocations (under NSF awards OCI-0725070, OCI 0832606 and ACI-1238993) and TACC's Frontera supercomputer allocations (PHY-20010 and AST-20021). Additional resources were provided by the RIT's BlueSky and Green Pairie and Lagoon Clusters 
acquired with NSF grants PHY-2018420, PHY-0722703, PHY-1229173 and PHY-1726215.

This work was performed in part at Aspen Center for Physics, which is
supported by NSF grant PHY-1607611, and at the Kavli Institute for Theoretical 
Physics, supported by the National Science Foundation under Grant No. NSF PHY-1748958.

Software: {\sc Harm3D} \citep{GMT03,Noble06,Noble09,Noble12}, {\sc Matplotlib} \citep{hunter2007Matplotlib}, {\sc NumPy} \citep{harris2020Array}, {\sc SciPy} \citep{virtanen2020SciPy}, and {\sc hdf5} \citep{hdf5}.

\end{acknowledgments}

\appendix
\section{Updates of PatchworkMHD}
\label{app:pwmhd}

As briefly summarized in Sec.~\ref{sec:setup}, the version of the Patchwork system used in this work is updated relative to the prior form \citep{Shiokawa2018} in several ways.  There have been a number of bug-fixes, efficiency-improvements, etc.   Most relevant to this study, we have also implemented magnetic field evolution and a system for suppressing magnetic field divergence on patch boundaries.  All these changes will be described in detail, including systematic testing for the new algorithms, in a dedicated code paper, Avara et al. (2024), though many are already summarized in \cite{Bowen2020}, which adapted the new PatchworkMHD code reported here to the numerical relativity context and reported on evolution of the wave equation reduction of Einstein's field equations. 

In summary, the improvements between the original Patchwork and PatchworkMHD can broadly be broken down into the following:

\begin{itemize}

	\item Significant rewrite of the multi-patch parallelization protocol to achieve scaling necessary to make evolution of astrophysical problems of the size of PM.IN20s affordable in HD and MHD. We improved on the interplay between local and global MPI comms, and removed communication bottlenecks, for instance the order of MPI send/recvs, to significantly reduce communication wait time and improve the scaling of the PatchworkMHD MPI parallelization with core count.
	\item The original Patchwork code function calls \citep{Shiokawa2018} were adjusted and implemented into the full Harm3D master branch, with accommodations made for utilization of any of the grid designs and space-times available in Harm3D.
	\item Harm3D core code was adjusted where necessary to accommodate the multi-patch infrastructure.
	\item Bug fixes, most notably related to memory leaks and ghost zone protocols utilized by MPI/multi-patch parallelization.
    \item Fundamental changes to how inter-patch zones are identified, creation of an optional buffer region, and a new hierarchical boundary condition structure to deal with patch boundaries that span both physical and inter-patch evolved regions.
    \item Addition of the magnetic field primitives to the communication infrastructure. 
    \item Changes to inter-patch interpolation and coordinate transformations that improve efficiency and accuracy.
    \item ``Sweeping" and ``untethering" algorithms to handle magnetic magnetic constrained transport, e.g. FluxCT, in the multipatch context
\end{itemize} 

We will now focus on the developments related to the last item of the list, specifically the untethering algorithm. Interpolation of magnetic field data from one patch to another when defining interpatch boundary conditions automatically introduces magnetic field divergence on the receiving domain, at the level of truncation error in the interpolation.  Successful evolution of MHD problems demands that this divergence be kept at a low level, the no-monopoles constraint.  We do so by two mechanisms, one needed even for patches with no relative motion (untethering), and the other additionally needed when patches move with respect to one another (sweeping).

The goal of untethering is to suppress growth in the magnetic divergence attached to patch boundaries due to the interpolation error when patches exchange boundary data.  The mechanism for doing so is to adjust the time-evolution of the magnetic field at the centers of a patch's outermost physical cells by adding small changes to the electromotive forces in a way that preserves zero-divergence within the problem domain while causing magnetic divergence on the patch boundary to decay.  In this fashion, it ``untethers" the open magnetic field loops associated with magnetic divergence at loop base points from the patch boundary.

We implement it as a series of  three steps.  In the first, the cells with faces lying on the physical boundary of a patch are identified.   For each of these cell-faces, we average the magnetic divergence on its four corners; this quantity is denoted by $D$.
In the second, we construct a pseudo-electric field associated with divergence, one designed to be non-zero only when there is inflow across the boundary.  We define this pseudo-electric field as
\begin{equation}
E^\prime = fC D \Delta x^1 u^1.
\end{equation}
Here the 1-direction is normal to the boundary, while the 2- and 3-directions are in the plane tangent to the boundary.  The cell dimension and fluid 4-velocity in the 1-direction are $\Delta x^1$ and $u^1$.  The coefficient $C$ is an adjustable parameter (typically $\approx 0.01$) whose value is the smallest capable of adequately controlling divergence growth. The factor $f$ introduces a sign-choice and an order-unity factor that depend on the direction of $u^1$ (inflow or outflow from the patch) and its magnitude relative to $|u^2|$ and $|u^3|$.
Defined in this way, $E^\prime$ is proportional to the contribution toward the numerical electric field associated with the magnetic divergence.

In the third step, we adjust the electric fields entering the Faraday equation update for the outermost physical cells by subtracting $E^\prime$ from the field components relevant to the patch-boundary surface:
\begin{equation}
\partial_t \left( \sqrt{-g} B^k\right) = - \partial_i \left[\sqrt{-g} \left(b^i u^k - b^k u^i\right) - E^\prime \delta_1^i)\right].
\end{equation}
In other words, the electric field whose curl changes the magnetic field at the centers of the
affected boundary cells is altered, but the FluxCT algorithm still preserves the (zero) magnetic divergence at those corners located on the FluxCT evolved domain, preventing divergence leakage into that domain, because the adjusted electric fields enter the calculation in the same way as the electric fields usually do, i.e., in loop-integration along cell edges.  However, this alteration diminishes the magnetic divergence on their outer corners by reducing the difference between the normal component of the magnetic field in the first ghost/interpatch cell and in the last physical cell, thereby bringing the solution closer to physical validity.

When patches move, magnetic divergence enters in a different way. Although this situation does not arise in the simulations reported here, we include this discussion for completeness.  In this case, the physical volume covered by different patches may overlap in a way that is time-dependent.  Where they overlap, only one of the patches controls the evolution, but motion causes these overlap regions to change.  When a cell in a certain patch becomes ``uncovered", that is, responsibility for evolution of its volume has been given to that patch, its fluid primitive variables can be safely interpolated from the data in the cells of the patch formerly responsible.  However, interpolation of magnetic field introduces divergence that must be eliminated.

To remove this sort of divergence, we use a new sweeping method.  Its basis is the natural desire to find a new value of the magnetic field at each uncovered cell-center such that the accumulated magnetic divergence from the distribution of interpolation error is transported to the edge of the evolved domain (and then damped by the untethering routine), and to do so in a way that minimally changes the field from the one found by interpolation.  The zero-divergence condition can be written down formally as the linear equation
\begin{equation}
C_{ij}X^j = D_i,
\end{equation}
where each value of $i$ labels a particular cell-corner where the divergence is evaluated, $X_j$ is a vector incorporating all the to-be-determined magnetic field components at all the relevant locations, and $D_i$ contains the summed information at each cell-corner stemming from fixed magnetic field components.  If there are $N$ cell-corners and $M$ unknown magnetic field components, typically $N < M$ because each cell-center has three field components.  In other words, this is an underdetermined system.

However, we also have a constraint that the values $X^j$ should be as close as possible to the candidate values $Y^j$ furnished by interpolation.  The combination of this constraint with the underdetermined linear problem has an optimal solution (see \cite{strang09} and the lecture notes for the course Introduction to Linear Dynamical Systems, Stephen Boyd, Stanford University):
\begin{equation}
\vec X = \left[\vec Y - (C C^T)^{-1} \,(C \vec Y - \vec D)\right].
\end{equation}
It is optimal in the sense that $||\vec X - \vec Y||$ is minimal.

The only  numerical complication posed by this method is calculation of the matrix inverse $(C C^T)^{-1}$.  We do so by the Gauss-Jordan method.

\hfill \break
\bibliography{bhm_references,references,bhm_references_new}

\end{document}